\documentclass[aps,superscriptaddress,showpacs,nofootinbib,eqsecnum]{revtex4}

\usepackage{amssymb}  
\usepackage{amsmath}  
\usepackage{natbib}
\usepackage{epsfig}  
\usepackage{graphicx}  
\usepackage{dcolumn}
\usepackage{bm} 
\usepackage[english]{babel}  
\usepackage[latin1]{inputenc}

\begin{document}

\title{Stochastic  Langevin equations: Markovian and non-Markovian dynamics}

\author{R. L. S.  Farias} \email{ricardo@dft.if.uerj.br}
\affiliation{Departamento de
  Ci\^{e}ncias Naturais, Universidade Federal de S\~{a}o Jo\~{a}o del
  Rei,\\ 36301-160, S\~{a}o Jo\~{a}o del Rei, MG, Brazil}
\affiliation{Departamento de F\'{\i}sica Te\'orica, Universidade do Estado do
  Rio de Janeiro, 20550-013 Rio de Janeiro, RJ, Brazil}

\author{Rudnei O.  Ramos} \email{rudnei@uerj.br} \affiliation{Departamento de
  F\'{\i}sica Te\'orica, Universidade do Estado do Rio de Janeiro, 20550-013
  Rio de Janeiro, RJ, Brazil}

\author{L. A. da Silva} \email{las.leandro@gmail.com}
\affiliation{Departamento de F\'{\i}sica Te\'orica, Universidade do Estado do
  Rio de Janeiro, 20550-013 Rio de Janeiro, RJ, Brazil}

\begin{abstract}

Non-Markovian stochastic  Langevin-like equations of motion are compared to
their  corresponding Markovian (local) approximations. The validity of the
local approximation for these equations, when contrasted with the fully
nonlocal  ones, is analyzed in details. The conditions for when the equation
in a local form can be considered a good approximation are then explicitly
specified. We study both the cases of additive and multiplicative noises,
including system dependent dissipation terms, according to the
{}Fluctuation-Dissipation theorem. 

\bigskip\bigskip
\centerline{\bf Published: Physical Review E80, 031143 (2009)}
\bigskip

\end{abstract}

\pacs{02.60.Cb, 05.10.Gg, 05.40.Ca}

\maketitle

\section{Introduction}
\label{sec1}

Most systems in nature cannot be regarded as purely closed systems but open
ones, interacting with an environment (e.g. a thermal bath).  The
evolution of these systems must then be non unitary, {\it i.e.}, interactions
with the environment must lead to dissipation as well stochastic effects,
which is the way the environment backreacts on the system. One common way for
describing this non unitary evolution is by means of stochastic Langevin-like
equations of motion. These non-deterministic equations of motion are used in
many systems of interest, such as simulating Brownian motion in (classical and
quantum) statistical mechanics and in other areas of physical
interest~\cite{reviews}.

A classical example of a system whose dynamics is modeled by a Langevin
equation of motion is the one that describes the Brownian motion of a
classical particle of coordinate $q$, unitary mass and subjected to a
potential $V(q)$ (as usual dots mean derivative with respect to time and
$V^{\prime}[q(t)] \equiv dV/dq$),

\begin{equation}
\ddot{q}(t) + \eta \dot{q}(t) + V^{\prime}[q(t)] = \xi(t)\;,
\label{langevin}
\end{equation}
where $\eta$ is a Markovian (local) dissipation term and $\xi(t)$ is a
stochastic term with white noise and Gaussian properties, satisfying
(throughout this work we consider the Boltzmann constant $k_B=1$)

\begin{equation}
\langle\xi(t)\rangle=0 \;,\;\;\; \langle\xi(t)\xi(t')\rangle = 2 T \eta
\delta(t-t')\;.
\label{white}
\end{equation}

Approaches with Langevin equations such as Eq.~(\ref{langevin}) and its
generalizations, are used in different contexts, e.g.  in classical
statistical mechanics to study problems with dissipation and noise, to
determine how order parameters equilibrate and in the studies such as dynamic
scaling and dynamic critical phenomena~\cite{critical,critical2}.

Though extensively used, equations of the form of Eq.~(\ref{langevin}),
with noise properties as given by Eq.~(\ref{white}), can only be considered
phenomenologically. This is because it implicitly assumes that the environment
interacts instantaneously with the system. This is a physically unacceptable
situation that violates causality, since the environment bath has no memory
time. Its is worth mentioning that a similar situation happens with the
description of the dynamics of a conserved order parameter by a Cahn-Hilliard
equation~\cite{CH}, which lacks causality~\cite{Jou1}. In the case of the
Cahn-Hilliard equation this  happens because, since it is a diffusion-reaction
type of equation, it should be characterized by microscopic scattering events.
In real systems scattering events proceed through finite time intervals,
which, consequently, must lead to finite memory effects. In order to fix this
bad behavior of the Cahn-Hilliard  equation, memory effects must be taken into
account, as explicitly shown recently in Ref.~\cite{KKR}. In a microscopic
description of the effects of the environment degrees of freedom on  some
select variables taken as the system, the same is expected to
happen. Dissipation and stochastic (noise) terms are expected to originate
from scattering events, thus giving origin to finite interaction times, that
reflect in the system's equation of motion as nonlocal ({\it i.e.}
non-Markovian) terms with memory effects. The simplest archetype  of this is
the description of the system-environment as being modeled by linearly coupled
harmonic oscillators \cite{oscillators}  (for a general review, see
e.g. Ref.~\cite{weiss}), which also become to be known as Caldeira-Leggett
type of models~\cite{caldeira}. The derived equation of motion for the system
variable, when the bath degrees of freedom are integrated out, leads to a
generalized Langevin  equation (GLE) of the form

\begin{eqnarray}
\ddot{q}(t)   +  \int_{t_0}^{t}   dt' D  (t-t')\, \dot{q}(t')+ V^{'}[q(t)] =
\xi (t)\;,
	\label{genlangevin}
\end{eqnarray}
where $t_0$ is some initial time, $D(t-t')$ is a dissipation kernel and the
noise term $\xi(t)$ is still Gaussian with zero mean but colored, {\it i.e.},
with two-point correlation, according to the Fluctuation-Dissipation theorem
for classical systems, and, thus $\xi(t)$ satisfies

\begin{eqnarray}
\langle\xi(t)\rangle=0 \;,\;\;\; \left\langle \xi (t)\xi (t')\right\rangle= T
D(t-t') \;,
\label{genxixi}
\end{eqnarray}
where averages are assumed to be taken with respect to a bath of free
oscillators with equilibrium distribution $\rho_B^{(0)}$ at some temperature
$T$.  Similar derivations in the context of field theory models (see, for
instance Refs.~\cite{GR,BMR}) also show the emergence of generalized equations
of motion of the form of Eq.~(\ref{genlangevin}), with a more complicated
structure depending on the form of the coupling of the system (e.g. some field
we are interested in the dynamics) with the environment (or bath fields, made
of the remaining fields other than the one representing the system field).

Despite the fundamental differences between Eqs.~(\ref{langevin}) and
(\ref{genlangevin}), we expect that when the time scale of the memory kernel
$D(t-t')$ is much smaller than any other time scales of the system,  the local
approximation (\ref{langevin}), with dissipation coefficient defined
by~\cite{ingold,hanggi}

\begin{equation}
\eta = \int_{-\infty}^{t} dt' D(t-t')\;,
\label{eta}
\end{equation}
can still be a good approximation for the system's dynamics and the effects of
finite memory  be negligible. There is also an immense saving of effort as
well as much more transparent understanding of the physics from a local
equation as opposed to a nonlocal one, since the former can generally be
analyzed with much less numerical treatment than the latter, thus it is a very
important question to know when, and how accurately,  a generally nonlocal
equation can be approximated by a local form.   Likewise, when the memory
constrains cannot be ignored, such as when the typical microscopic time scales
are large in comparison to the other time scales characterizing the
dynamics, we must then be able to have appropriate tools to tackle the
non-Markovian equations of motion. This is possible with some restrict forms
of kernels, as we are going to see below, which, nevertheless, can represent
physically relevant systems of interest, thus, being well motivated.

In this work our objective is to gauge the applicability of an equation of
motion of the form of Eq.~(\ref{langevin}) when compared to the
non-Markovian form, when considering some of the most common forms for  the
dissipation kernel $D(t,t')$. These forms for the dissipation kernel include
for example the one that describes an Ornstein-Uhlenbeck (OU)
process~\cite{OUref} and the  exponential damped harmonic (EDH)
kernel~\cite{EDHref} (see also Ref.~\cite{Luczka} for a recent review on the
different colored noise terms used in the literature).   We study both the
cases of additive and multiplicative  noises, including system dependent
dissipation terms, according to the {}Fluctuation-Dissipation theorem.  A
detailed numerical analysis is made when the various parameters characterizing
the thermal bath, e.g. the bath relaxation (or damping) parameter, frequency and
temperature of the bath are varied. 

The remaining of this work is organized as follows. In Sec. II we  define the
prescription to transform the non-Markovian equations in a system  of
Markovian time differential equations. We study specifically the OU and EDH
kernels.  In Sec. III we present the numerical results for the non-Markovian
equations for different model parameters and compare the results to those
coming from their Markovian approximations.  The results are obtained for both
the cases of additive and multiplicative noises.  How the dynamics depends on
the various parameters characterizing the thermal bath is studied in
details. In Sec. IV we present our conclusions, discuss the various results we
have obtained,  and we give a possible relation and the relevance of our
results for the study of nonequilibrium dynamics in field theory models.


\section{The non-Markovian Langevin-like Equation}
\label{sec2}

Here we study a GLE describing the interaction of a system, denoted by  a
variable $\phi$ (which can be e.g.  the coordinate of a particle) 
in interaction with a
thermal bath, where the noise has the properties such as in Eq.~(\ref{genxixi}).
The GLE studied here is then of the generic form

\begin{eqnarray}
  \ddot{\phi}(t)   +  \phi^n(t) \int_{t_0}^{t}   dt'  \phi^n(t') D (t-t')\,
  \dot{\phi}(t')+ V'(\phi) = \phi^n(t)\, \xi (t)\;,
  \label{gle}
\end{eqnarray}
where $n=0,1$, with $n=0$ giving the standard GLE of additive form, such as
Eq.~(\ref{genlangevin}), while for $n=1$ gives a multiplicative GLE. The
multiplicative noise and system dependent dissipation 
form are motivated from field theory calculations \cite{GR,BMR} and this is 
why we have also 
included this special case here in view of future applications in that
case. The potential in Eq.~(\ref{gle}) is considered to be one with quadratic
and quartic terms, given by

\begin{equation}
V(\phi) = \frac{m^2}{2}\phi^2  + \frac{\lambda}{4!}\phi^4\;,
\label{pot}
\end{equation}
where $m^2$ and $\lambda$ are parameters depending on the details of the
system under study. Here, we can associate $m$ with the system's frequency and
$\lambda$ with the degree of nonlinearity of the system's potential (or in the
context of field theories, with the strength of the system's
self-interactions). 

Nonlinear GLEs of the form of Eq.~(\ref{gle}) are notably difficult to solve.
Analytical methods can only be used when the equation can be approximated  or
put in a linear form, such as in the additive noise case and when the quartic
term in the system's potential can be neglected. 
This is because, in the additive noise and variable
independent dissipation case, such as in Eq.~(\ref{genlangevin}), the equation is
in the form of a convolution, so can be solved through Laplace transform for
instance~\cite{CPC2008}. Otherwise, in the more general cases,  we must resort
to numerical methods. This is the approach we follow in this paper in order to
analyze the dynamics obtained from Eq.~(\ref{gle}). 
Though there are some
specific numerical methods using e.g. {}Fourier transform that may apply for
equations with non-Markovian kernels of generic form~\cite{bao1}, we still
would like to be able to solve equations such as Eq.~(\ref{gle}) through standard
methods, which are less numerically expensive than other alternatives. This is
the case, for example, when using Runge-Kutta methods. Recently, the
authors~\cite{CPC2008}  have demonstrated the reliability of using a
fourth-order Runge-Kutta method  when solving GLE of the OU and EDH forms. The
way this can be done stems from the fact that non-Markovian equations with
kernels of those forms can be replaced by a system of completely local
first-order differential equations, which has been described in details
in~\cite{CPC2008}.

As already mentioned, in this work we concentrate our study in equations  such as
Eq.~(\ref{gle}) with non-Markovian kernels of either the OU type \cite{OUref}

\begin{equation}
D_{OU}(t-t') =  \eta\, \gamma e^{-\gamma (t-t^{\prime })}\;,
\label{DOU}
\end{equation}
or with the EDH type \cite{EDHref}

\begin{eqnarray}
D_H(t-t^{\prime })&=&\eta \, e^{-\gamma(t-t^{\prime })} \frac{\Omega _{0}^{2}
}{2 \gamma }\left\{ \cos [\Omega _{1}(t-t^{\prime })] \right. \nonumber \\
&+& \left. \frac{\gamma }{\Omega_{1}}\sin [\Omega _{1}(t-t^{\prime })]\right\} \;,
\label{DEDH}
\end{eqnarray}
where in the equations above, $\eta$ sets the magnitude of the dissipation,
$\gamma$ sets the relaxation time for the bath kernels, $\tau = 1/\gamma$, and
$\Omega_0$ gives the oscillation time scale in the case of the EDH kernel.  In
Eq.~(\ref{DEDH}), $\Omega_1^2 =\Omega _{0}^{2}-\gamma^{2}$, and so, 
in the EDH case the values of $\gamma$ and $\Omega_0$ are restricted so to have
$\Omega_1^2\geq 0$.

It can be easily shown that the OU and EDH noises can be generated by the
stationary part of the solution of the following differential equations,
respectively,

\begin{equation}
\dot{\xi}_{OU}(t) = -\gamma\left[\xi_{OU}(t) - \sqrt{2T \eta}\,
  \zeta(t) \right]\;,
\label{xiOUeom}
\end{equation}

\begin{equation}
  \ddot{\xi}_H(t)+ 2 \gamma \dot{\xi}_H(t)+\Omega_{0}^{2}\xi_H (t)=
  \Omega_{0}^{2}  \sqrt{2T \eta}\,\zeta(t)\;,
  \label{xiEDHeom}
\end{equation}

\noindent
where $\zeta$ in Eqs.~(\ref{xiOUeom}) and (\ref{xiEDHeom}) is a white Gaussian
noise satisfying

\begin{eqnarray}
\langle \zeta(t) \rangle &=&0\;,  \nonumber \\ \langle \zeta(t)
\zeta(t^{\prime }) \rangle &=& \delta(t-t^{\prime})\;.
\label{zeta}
\end{eqnarray}

Taking $t_0=0$, we can define a new variable $w(t)$ by~\cite{bao2,CPC2008}

\begin{equation}
w(t) = - \int_{0}^{t} dt' \phi^n(t') D (t-t')\dot{\phi}(t')\;.
\label{wt}
\end{equation}
This, together with Eqs.~(\ref{xiOUeom}) and (\ref{xiEDHeom}), leads to the
following system of local first-order differential equations representing the
GLE Eq.~(\ref{gle}): {}For the OU case,

\begin{eqnarray}
\dot{\phi} &=& y\;,  \nonumber \\ \dot{y} &=& - V^{\prime}(\phi) + \phi^n
w_{OU} + \phi^n \xi_{OU} \;, \nonumber \\ \dot{w}_{OU} &=& -\gamma w_{OU}-
D_{OU}(0) \phi^n y \;, \nonumber \\ \dot{\xi}_{OU} &=& -\gamma\left[\xi_{OU} -
  \sqrt{2T \eta}\,\, \zeta\right]\;,
\label{setOU}
\end{eqnarray}
while for the EDH case we obtain

\begin{eqnarray}
  \dot{\phi}&=&y\;,  \nonumber \\ \dot{y}&=&-V^{\prime }(\phi )+ \phi^n w_{H}
  + \phi^n \xi_H \;,  \nonumber \\ \dot{w}_{H}&=&u - 2 \gamma w_{H} - D_H(0)\,
  \phi^n y\;,  \nonumber \\ \dot{u}&=&-\Omega_{0}^{2} w_{H} +\dot{D}_{H}(0)\,
  \phi^n y -  2 \gamma D_{H}(0)\, \phi^n y\;, \nonumber
  \\ \dot{\xi}_{H}&=&z\;,  \nonumber \\ \dot{z}&=&-2\gamma
  z-\Omega_{0}^{2}\xi_{H} + \Omega_{0}^{2}\sqrt{2T \eta}\, \zeta\;,
\label{setEDH}
\end{eqnarray}
where the variable $u(t)$ in Eq. (\ref{setEDH}) is defined as

\begin{equation}
u(t) = \int_{0}^t dt^{\prime }\left[ \frac{d D_H(t-t^{\prime })}{dt'} -
2 \gamma D_H(t-t^{\prime }) \right] \phi^n(t') \frac{d\phi(t^{\prime })}{dt'}\;,
\end{equation}
and in Eqs. (\ref{setOU}) and (\ref{setEDH}), we also have that
$D_{OU}(0) = \eta\,\gamma$, $D_H(0)=\eta \,
\Omega_0^2/(2\gamma)$ and $\dot{D}_{H}(0) = 0$,   which follow from
Eqs.~(\ref{DOU}) and (\ref{DEDH}).  The additive noise case is when $n=0$ is
taken in Eqs.~(\ref{setOU}) and (\ref{setEDH}), while $n=1$ is for the
multiplicative noise case.  These two cases, for both types of kernels, are
next numerically studied below.

\section{Contrasting the non-Markovian numerical results with the
Markovian approximation ones}

Let us now consider our numerical results for the Markovian and non-Markovian
dynamics for the system. In the Markovian approximation all memory effects are
neglected and the non-Markovian dissipation term in Eq.~(\ref{gle}) is
replaced by a local dissipation term with magnitude as given by
Eq.~(\ref{eta}), {\it i.e.}, we write Eq.~(\ref{gle}) in the form

\begin{eqnarray}
  \ddot{\phi}(t)   +  \eta \, \phi^{2n}(t)\, \dot{\phi}(t)+ V'(\phi) =
  \phi^n(t) \,  \xi (t)\;.
  \label{glelocal}
\end{eqnarray}

\noindent
As we have mentioned before, in general we expect the local form the GLE to be a
valid approximation when the relaxation time scale for the thermal bath, $\tau
= 1/\gamma$, is much smaller than the characteristic time scale for the
system,  e.g., $\tau \ll \phi/\dot{\phi}$ (this is equivalent to the
quasi-adiabatic condition set in Ref.~\cite{GR} in the field theory case for
the validity of the local  Markovian approximation).  When this condition is
met in a sufficiently large time interval $\Delta t = t-t_0$, thus $\Delta
t/\tau \gg 1$ (which is equivalent as taking $t_0 \to -\infty$)  and the time
nonlocal term in Eq.~(\ref{gle}) can be written as

\begin{eqnarray}
&&\phi^n(t)\int_{t_0}^{t} dt' D(t-t') \phi^n(t')\, \dot{\phi}(t') 
\nonumber \\
&& \simeq \phi^{2n}(t)\, \dot{\phi}(t) 
\int_{t_0\to -\infty}^{t} dt' D(t-t')  \to 
\eta \, \phi^{2n}(t)\, \dot{\phi}(t)\;,
\label{localapprox}
\end{eqnarray}
where in the last step we have used the definition Eq. (\ref{eta}).  The
result (\ref{localapprox}) then leads to the local dissipation term  in
Eq.~(\ref{glelocal}).  Of course, under the conditions set above, at
sufficiently short times we  expect the memory effects to influence the
dynamics in some significant way, but these memory effects should quickly
become negligible at long times, $t \gg 1/\gamma$.  In any event, we thus
expect that after some long time period the memory effects can become
sufficiently damped such that the Markovian approximation could represent well
the overall dynamics of the system. After all, we expect that both dynamics,
the non-Markovian and the Markovian ones to both have the same asymptotic
state. But we still face with a natural and important question to answer:
{}For a given set of model parameters representing the system and the thermal
bath to which it is coupled to, for how long can we expect the memory effects
due to the non-Markovian terms to be important and when can they be neglected
and then the dynamics be well represented by the local Eq.
(\ref{glelocal})  ? This is because, even though both dynamics are expected to
approach each other asymptotically, the time this happens could be so long that
the memory effects could lead to important physical effects and the local
Markovian dynamics would just not be appropriate to be used.  Since the
representation of the dynamics in a local form as given by
Eq.~(\ref{glelocal}) represents a considerable simplification, for both a
numerical point of view, or for analytical analysis (when it  is possible),
when compared, e.g., to the full nonlocal, integro-differential stochastic
Eq. (\ref{gle}), this then becomes an important question to be accessed
for most practical studies  that make use of stochastic equations of
motion. It is also important to investigate how the dynamics is affected by
varying not only $\gamma$, but the other parameters characterizing the thermal
bath, such as the temperature $T$ and frequency $\Omega_0$, whose effects
on the dynamics may not so direct as the ones obtained by just varying
$\gamma$.
Below we try to answer all these questions, performing our study, numerically,
in the context of the additive and
multiplicative noise cases, with either the OU or EDH kernel terms, defined in
the previous section.

Next we show the results of our systematic simulations of the system of
differential first-order equations, Eqs.~(\ref{setOU}) and (\ref{setEDH}), for
the non-Markovian GLE with OU and EDH kernels, respectively. The results are
compared to those obtained through the local approximation given by
Eq.~(\ref{glelocal}). All our simulations were performed with $300\,000$
realizations over the noise and we have integrated all differential equations
using a standard fourth-order Runge-Kutta method with a  time stepsize varying
between $\delta t= 0.01$ and $\delta t= 0.001$, 
which were found to be more than enough for both numerical
stability and also for enough numerical precision  (as determined in
Ref. \cite{CPC2008}, these values already assure an overall numerical error of
always smaller than  about one percent, which suffices for our comparison
purposes set here). In all our simulations we have also used the initial
conditions $\phi(0)=1$ and $\dot\phi(0)=0$. The time in all our evolutions is
in units of the (inverse of the) frequency for the system (which is equivalent
to consider $m=1$ throughout).  Comparisons between the Markovian and
non-Markovian dynamics are made varying the relevant parameters of the bath
for the two cases of kernels considered, while keeping the system parameters
fixed.

\subsection{The additive noise case}

\begin{figure}[htb]
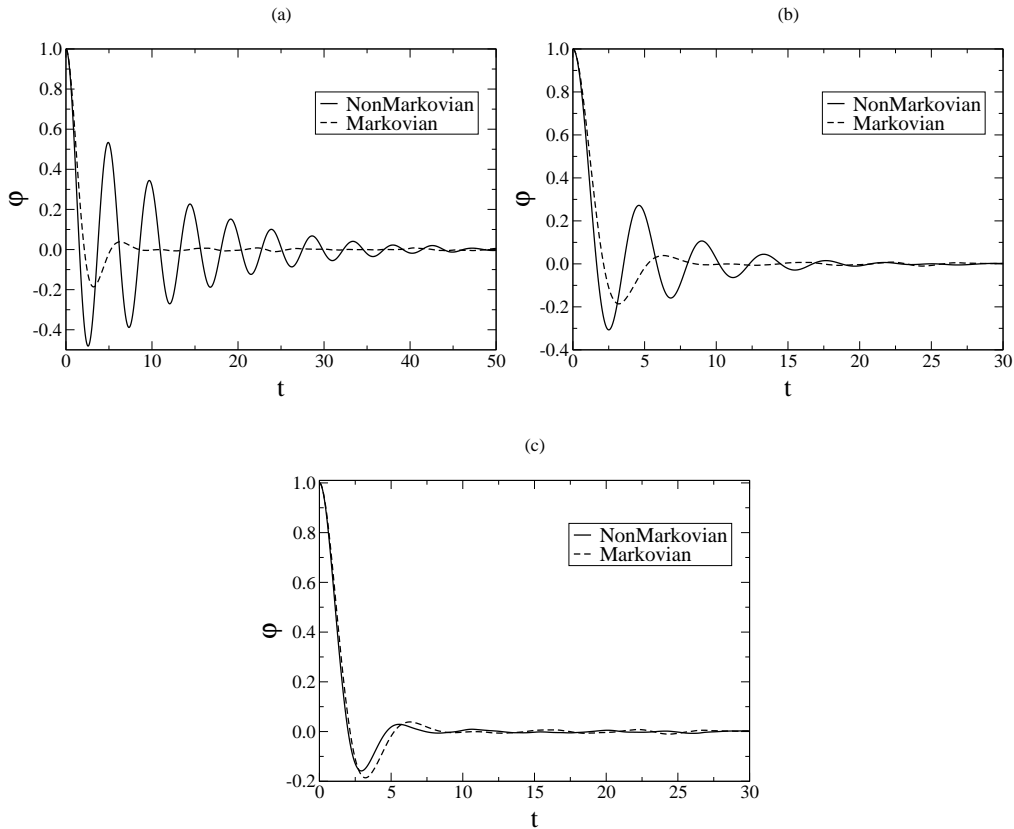

 \centerline{ \psfig{file=fig1a.eps,scale=0.265,angle=0}
   \psfig{file=fig1b.eps,scale=0.265,angle=0}}
\vspace{0.5 cm}
\centerline{ \psfig{file=fig1c.eps,scale=0.265,angle=0}}
   \caption{\sf OU case with additive noise and its Markovian approximation:
     time evolution for $\varphi(t)$. (a) $\gamma=0.5$, (b)  $\gamma=1.0$ and
     (c) $\gamma=5.0$. The other parameters are taken as $\Omega_0=1.0$,
     $\eta=1.0$, $T=1.0$, $m=1.0$ and $\lambda=1.0$.}
   \label{fig1}
\end{figure}


\begin{figure}[htb]
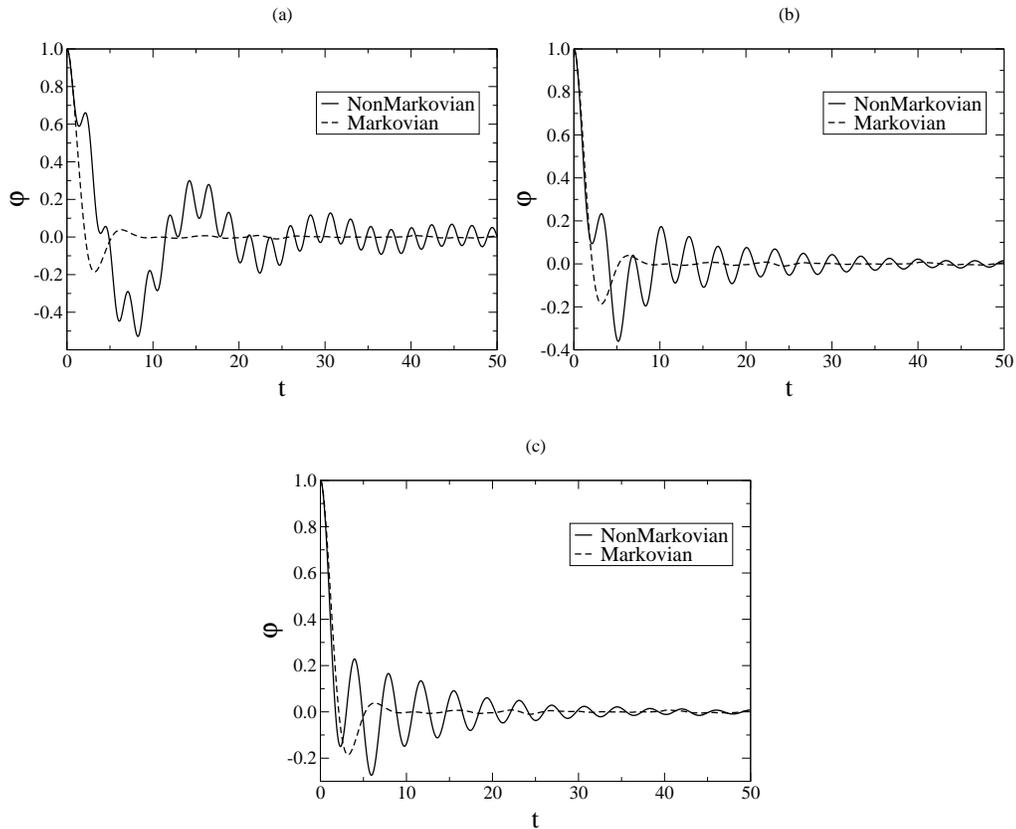

 \centerline{ \psfig{file=fig2a.eps,scale=0.265,angle=0}
   \psfig{file=fig2b.eps,scale=0.265,angle=0}}
\vspace{0.5 cm}
\centerline{ \psfig{file=fig2c.eps,scale=0.265,angle=0}}
   \caption{\sf EDH case with additive noise and the Markovian approximation:
     time evolution for $\varphi(t)$. (a) for $\gamma=0.1$, (b) for
     $\gamma=0.3$ and (c) for $\gamma=0.5$. The other parameters are the same
     as in {}Fig. 1.}
   \label{fig2}
\end{figure}


Let us now turn to our numerical results.   The relevant bath parameters are
the dissipation magnitude $\eta$, the  temperature $T$ (that are common to
both Markovian and non-Markovian dynamics), the bath damping parameter
$\gamma$, and the bath frequency  $\Omega_0$ (in the EDH kernel case). Since
the dissipation magnitude is common for both types of dynamics, in the
following we keep $\eta$ fixed, at the value $\eta=1$ throughout
(which can be checked
to correspond to lead to an underdamped dynamics in the local equations of
motion) and we vary the remaining bath parameters. This will
allow us to better understand the importance of the memory effects alone for the
dynamics. So, we consider the various dynamics as $\gamma$, $\Omega_0$ and $T$
are changed.  We include the study of how the dynamics changes with the
temperature  because this is useful to determine how this internal property of
the  thermal bath influences the dynamics when comparing to both the 
Markovian and non-Markovian cases.

We start our analysis by first considering variations in the  parameter
$\gamma$. Representative values for $\gamma$ are then  chosen and all the
remaining parameters are  initially kept fixed. Note that since $\gamma$ acts
as damping the effects of the nonlocal kernels, the larger is $\gamma$ the
better must be the local approximation for the full nonlocal dynamics. 
Then we keep $\gamma$ fixed at the largest value used in our analysis and
then consider variations in the other parameters.
This will allow us to determine
the consequent importance of the remaining parameters and whether a change of
those parameters can discriminate the types of dynamics studied, for example
discriminate additive and multiplicative stochastic dynamics.

Let us first consider the GLE with additive noise and OU kernel.  This is
considered in {}Fig. \ref{fig1}, where we plot side by side  our results for
the dynamics of the ensemble averaged  macroscopic system variable $\phi$,
$\langle \phi \rangle =\varphi(t)$, where the average is over the noise
realizations. This is obtained from the Markovian and non-Markovian equations,
Eqs.~(\ref{glelocal}) and (\ref{setOU}), respectively, with $n=0$.

In {}Fig. \ref{fig2} we plot side by side our results for $\varphi(t)$  for
the Markovian and non-Markovian regimes for the EDH case, again by considering
the  additive noise ($n=0$).

The effect of changing $\gamma$ seen in both {}Figs. \ref{fig1} and \ref{fig2}
is clear and well within the expected: The larger is the relaxation time
($1/\gamma$) for the nonlocal kernels, the larger are the memory effects,
resulting in a strong difference with respect to the local approximation, seen
most notably at short times.   As also expected, at some sufficient long time,
that we here see to depend on  how large $\gamma$ is, the two dynamics,
Markovian and non-Markovian approximate one of the other.  This is also seen
if we had plotted the correlation $\langle \phi^2(t) \rangle$ for both OU and
EDH cases.  The difference between the Markovian and non-Markovian dynamics
can also be better estimated by defining the quantities,

\begin{eqnarray}
\Delta \phi  &=& \langle \phi \rangle _{\rm non-Markovian} -  \langle
\phi\rangle_{\rm Markovian}\; \nonumber \\ \Delta\phi^2  &=& \langle \phi^2
\rangle_{\rm non-Markovian} -  \langle \phi^2 \rangle_{\rm Markovian}\;.
\label{Deltas}
\end{eqnarray}

\begin{figure}[htb]
\vspace{0.95 cm} \centerline{ \psfig{file=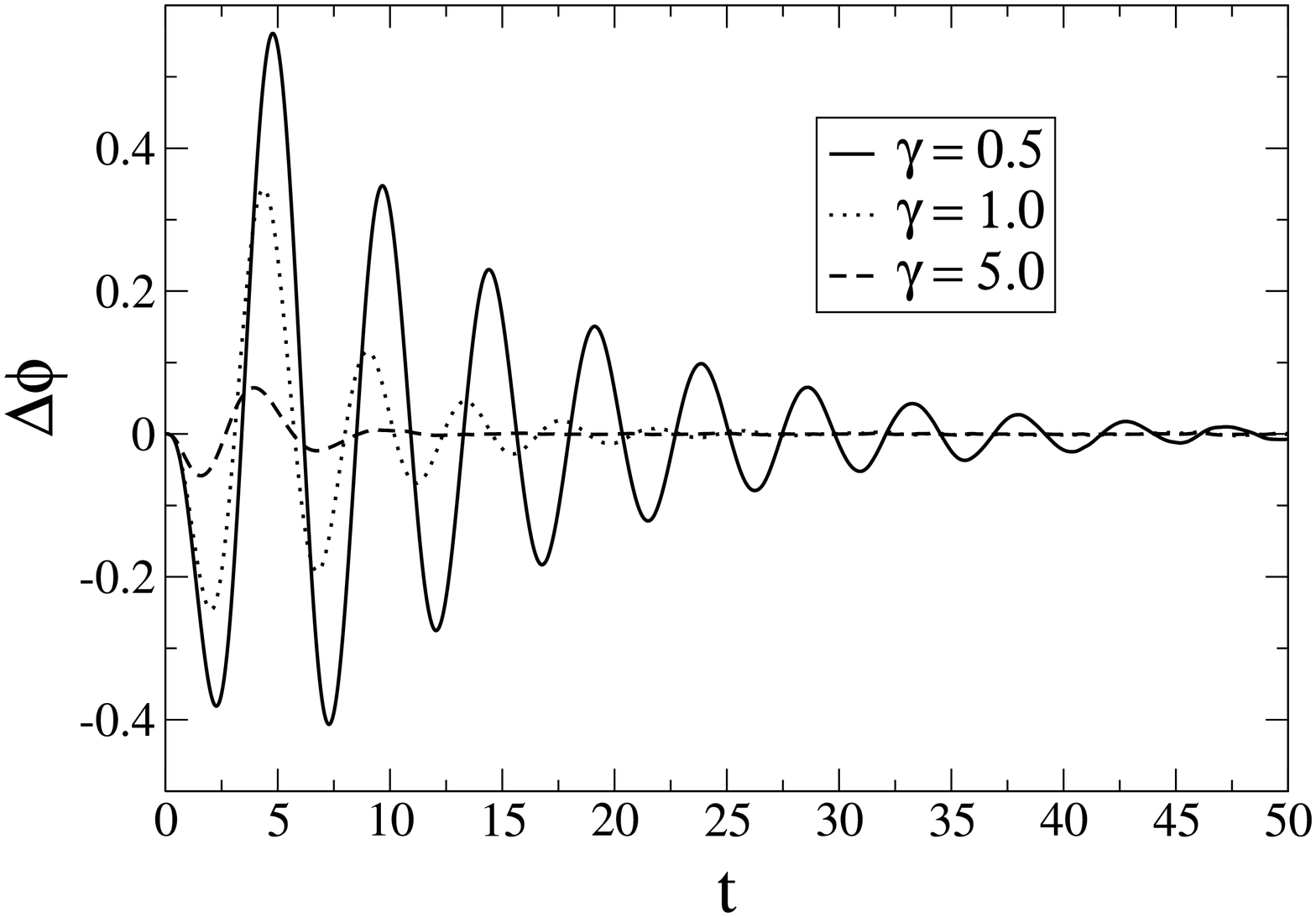,scale=0.275,angle=0}
  \psfig{file=fig3b.eps,scale=0.275,angle=0}}
  \caption{\sf The differences $\Delta \phi$ and $\Delta \phi^2$ in the OU
    additive case.  All other parameter kept fixed as before.}
  \label{fig3}
\end{figure}

\begin{figure}[htb]
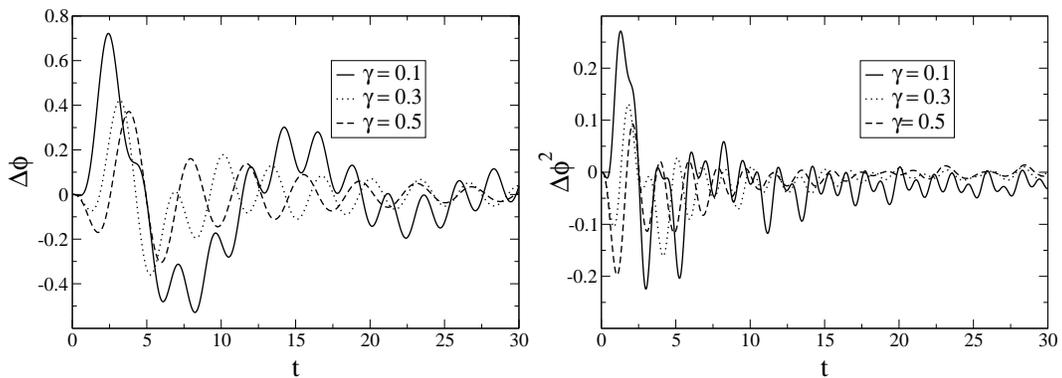

  \centerline{ \psfig{file=fig4a.eps,scale=0.275,angle=0}
    \psfig{file=fig4b.eps,scale=0.275,angle=0}}
  \caption{\sf The differences $\Delta \phi$ and $\Delta \phi^2$ in the
    difference $\Delta \phi^2$ in the EDH additive case.}
  \label{fig4}
\end{figure}

The results for the differences $\Delta \phi$ and $\Delta \phi^2$ are shown in
{}Figs. \ref{fig3} and \ref{fig4}, for the OU and EDH cases, respectively.

\noindent
The results from the plots shown in {}Figs. \ref{fig3} and \ref{fig4}  are
useful to determine within which time scale the Markovian and non-Markovian
dynamics become sufficiently close (within to some  given precision). The
results for these time scales for the different simulations we have performed
for the Markovian dynamics and for the non-Markovian dynamics with the two
types of memory kernels explored here will be given below in Table \ref{tab1}.  

An immediate conclusion we can realize from an inspection of the results  
shown in {}Figs. \ref{fig1} - \ref{fig4} is that the time scale that 
it takes for the
two dynamics to begin to be  equivalent is much larger than the time scale for
the kernel relaxation itself, $1/\gamma$, and also larger than the  typical
system's time scale, which is typically given by the inverse of  the system's
frequency ($1/m$). This will remain true for the multiplicative noise case with
either the OU or the EDH memory kernels.

\subsection{The multiplicative noise case}

\begin{figure}[htb]
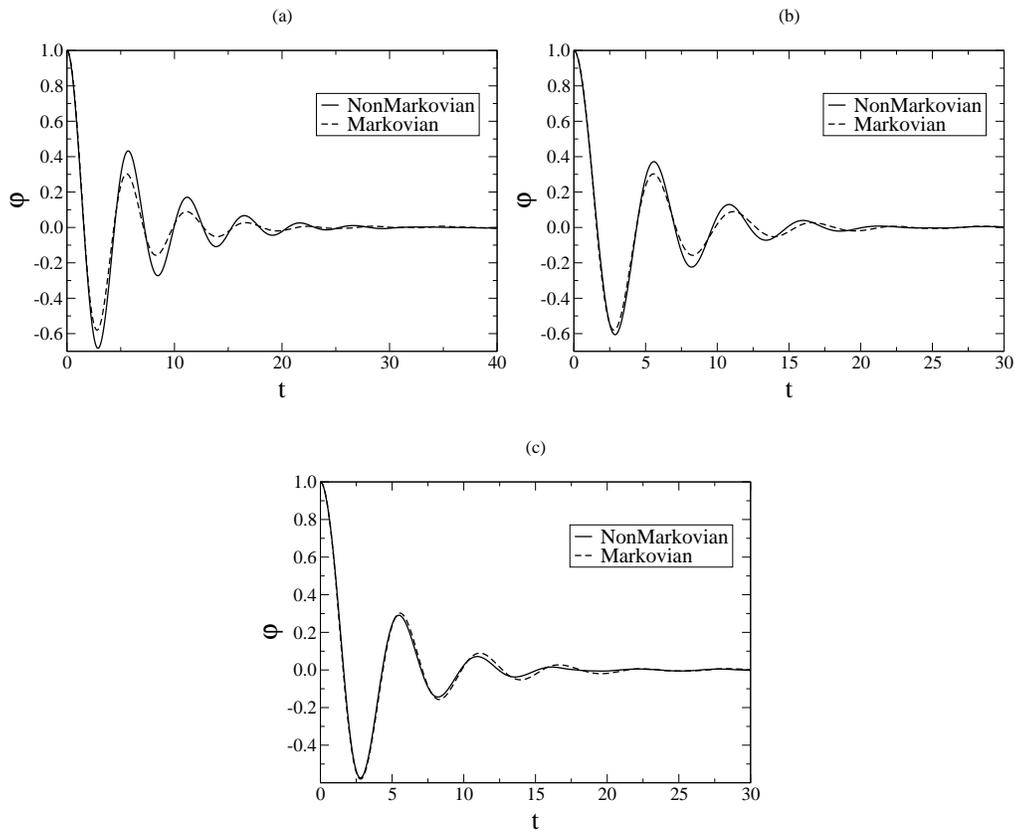

 \centerline{ \psfig{file=fig5a.eps,scale=0.265,angle=0}
   \psfig{file=fig5b.eps,scale=0.265,angle=0}}
\vspace{0.5 cm}
\centerline{ \psfig{file=fig5c.eps,scale=0.265,angle=0}}
   \caption{\sf OU case with multiplicative noise and its  Markovian
     approximation:  time evolution for $\varphi(t)$. (a) for $\gamma=0.5$,
     (b) for $\gamma=1.0$ and (c) for $\gamma=5.0$. The other parameters are
     taken as $\Omega_0=1.0$, $\eta=1.0$, $T=1.0$ and $\lambda=1.0$.}
   \label{fig5}
\end{figure}


\begin{figure}[htb]
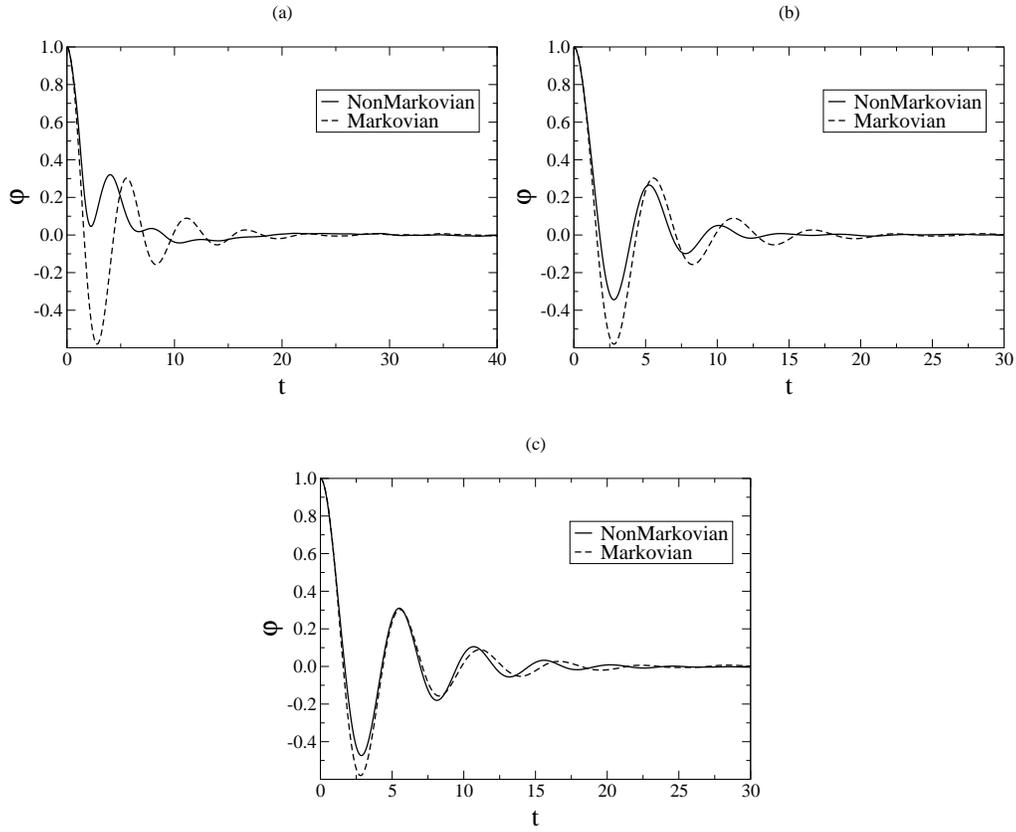

 \centerline{ \psfig{file=fig6a.eps,scale=0.265,angle=0}
   \psfig{file=fig6b.eps,scale=0.265,angle=0}}
\vspace{0.5 cm}
\centerline{ \psfig{file=fig6c.eps,scale=0.265,angle=0}}
   \caption{\sf EDH case with multiplicative noise and the  Markovian
     approximation:  time evolution for $\varphi(t)$.  (a) for $\gamma=0.1$,
     (b) for $\gamma=0.3$ and (c) for $\gamma=0.5$. All other parameters taken
     as in the previous figures.}
   \label{fig6}
\end{figure}


Let us now verify the results concerning the Markovian and non-Markovian
stochastic equations when multiplicative noise and system dependent
dissipation are concerned. We then here explore the case $n=1$ in
Eqs.~(\ref{setOU}) and (\ref{setEDH}), for the OU and EDH cases, respectively.

In {}Fig.~\ref{fig5} we plot side by side our results  for  $\langle \phi
\rangle$ for the Markovian and non-Markovian regimes in the OU case with
$n=1$, while in {}Fig.~\ref{fig6} are the results for the EDH case.

\begin{figure}[htb]
\vspace{0.95 cm} \centerline{ \psfig{file=fig7a.eps,scale=0.265,angle=0}
  \psfig{file=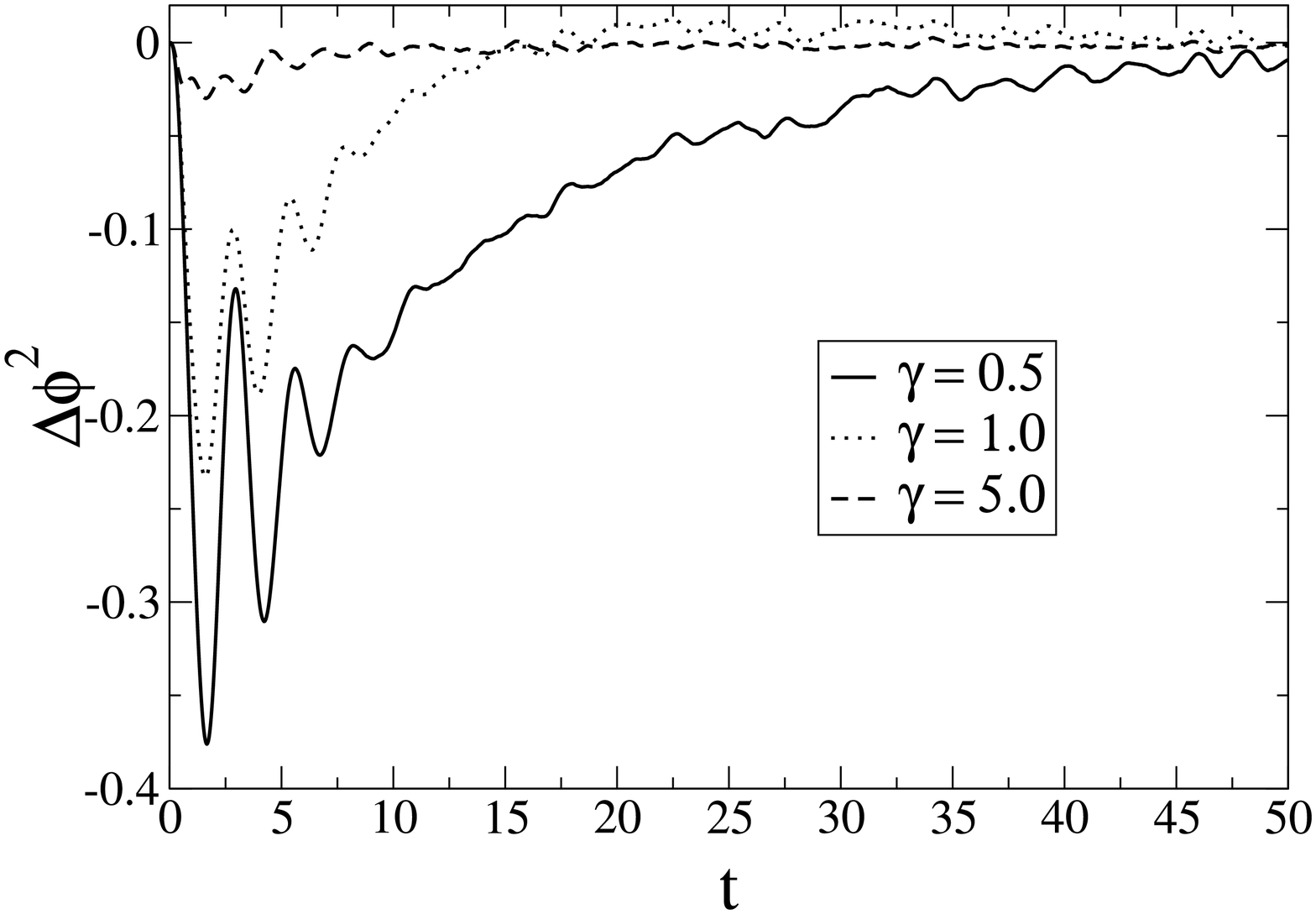,scale=0.265,angle=0}}
  \caption{\sf The differences $\Delta \phi$ and $\Delta \phi^2$ for the OU
    multiplicative case.  All other parameter kept fixed as before.}
  \label{fig7}
\end{figure}

\begin{figure}[htb]
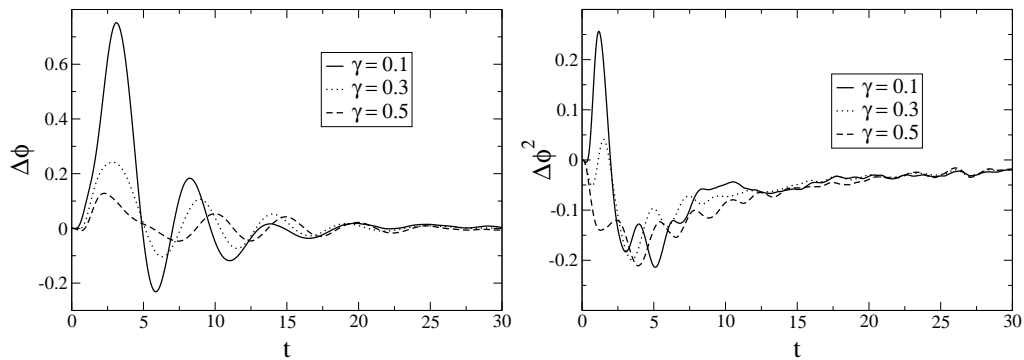

\centerline{ \psfig{file=fig8a.eps,scale=0.265,angle=0}
  \psfig{file=fig8b.eps,scale=0.265,angle=0}}
  \caption{\sf The differences $\Delta \phi$ and $\Delta \phi^2$ for the EDH
    multiplicative case.}
  \label{fig8}
\end{figure}

The results for the differences $\Delta \phi$ and $\Delta \phi^2$, for the
$n=1$ multiplicative noise case, are shown in {}Figs. \ref{fig7} and
\ref{fig8}, for the OU and EDH cases, respectively.

The results shown in {}Figs. \ref{fig5} - \ref{fig8} again indicate that, as
expected and similar to the additive noise  $n=0$ case, that the two dynamics,
local and nonlocal, become closer to each other the larger is the kernel
damping parameter. They also seem to indicate that in the multiplicative noise
case the two dynamics are closer to each other for the same parameters used in
the additive noise case. This can be more quantitatively estimated through the
differences (\ref{Deltas}). The results for the time scales when the two
dynamics begin to be sufficiently close to each other are also presented in
Table \ref{tab1}.

\begin{table}[pt]
{\begin{tabular}{@{}c|c|c|c|c@{}}  
\hline  
$\gamma$ & $\Delta\phi_{\rm
    OU \; add}$ & $\Delta\phi^2_{\rm OU \; add}$ &  $\Delta\phi_{\rm OU \; mult}$ &
$\Delta\phi^2_{\rm OU \; mult}$  \\ 
\colrule 
0.5  & 73  & 28  &  35  & 70 \\ 
1.0  & 32 & 25 & 30  & 24 \\ 
5.0  & 14 & 7  & 27 & 12 \\ 
\hline
\end{tabular}}
\hspace{0.3cm}
{\begin{tabular}{@{}c|c|c|c|c@{}}  
\hline  
$\gamma$ & $\Delta\phi_{\rm
    EDH \; add}$ & $\Delta\phi^2_{\rm EDH \; add}$ &  $\Delta\phi_{\rm EDH \; mult}$ &
$\Delta\phi^2_{\rm EDH \; mult}$  \\ 
\colrule 
0.1  & 180  & 89    & 30  & 73 \\ 
0.3  & 67   & 33    & 28  & 71 \\ 
0.5  & 53   & 26    & 23 & 67 \\ 
\hline 
\end{tabular}}

\caption{The approximate time scale, in units of $1/m$, for the non-Markovian
  dynamics to approach the Markovian one, within a precision of $10^{-3}$ for
  the differences defined in (\ref{Deltas}).}
\label{tab1}
\end{table}

We note from the results shown in Table \ref{tab1} that the system variable
$\phi$ in the case of the GLE with OU additive noise tends to approximate the
corresponding  Markovian dynamics faster than in the case of the
multiplicative noise case.  The opposite seems to happen to the equal time
correlation function  $\langle \phi^2 \rangle$, where in the case of OU
multiplicative noise it is faster than in the case of additive noise. 
The behavior of $\Delta\phi$ in the additive noise cases indicates that
the averaged system variable $\langle \phi \rangle$ is approached faster
to the Markovian dynamics  than in the multiplicative 
noise cases. The behavior for the dynamics of $\langle \phi^2
\rangle$, when looking at the behavior of $\Delta\phi^2$, is analogous,
except in the EDH multiplicative noise case, where it is slower than in the 
additive noise case.

\begin{figure}[htb]
 \centerline{ \psfig{file=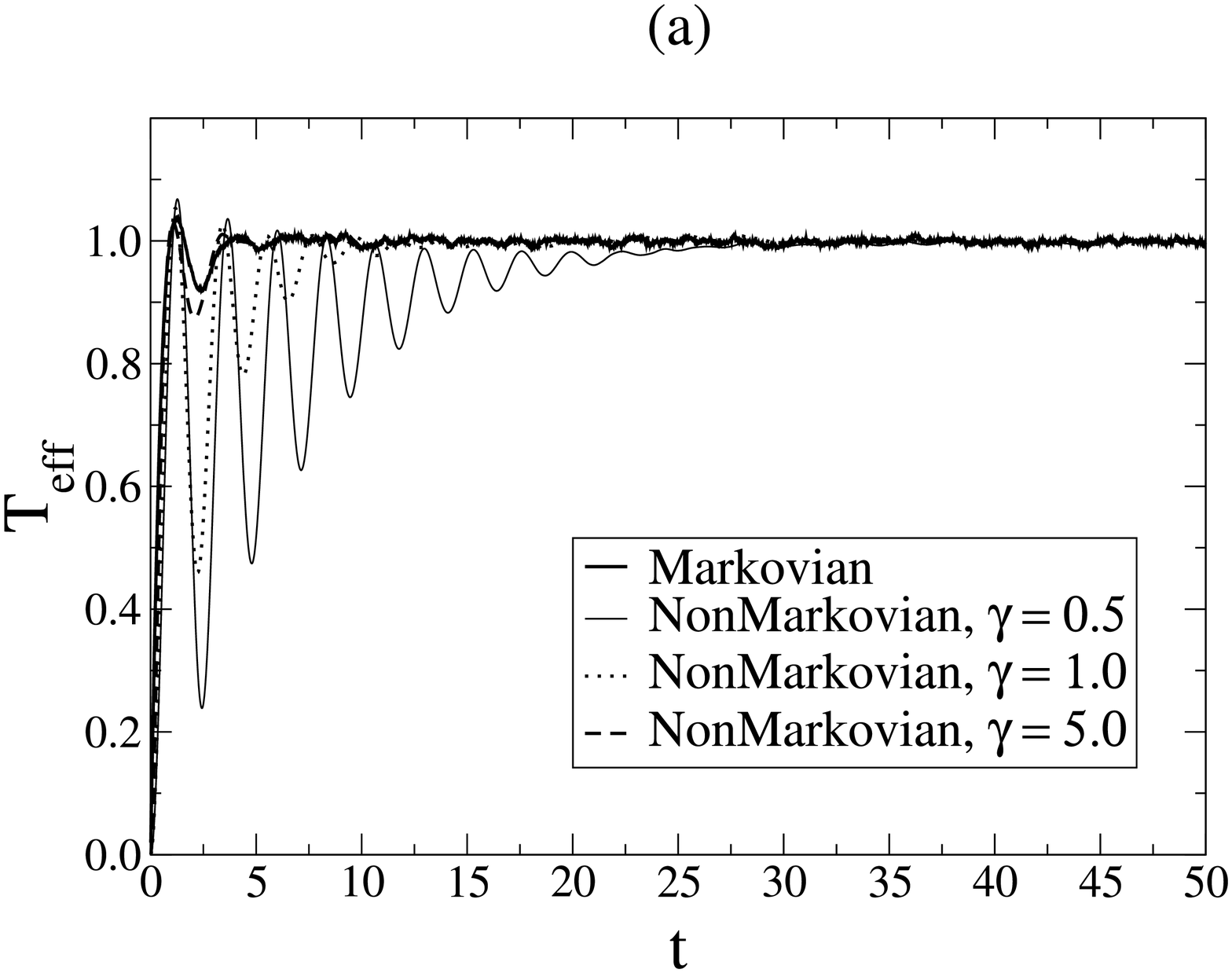,scale=0.265,angle=0}
   \psfig{file=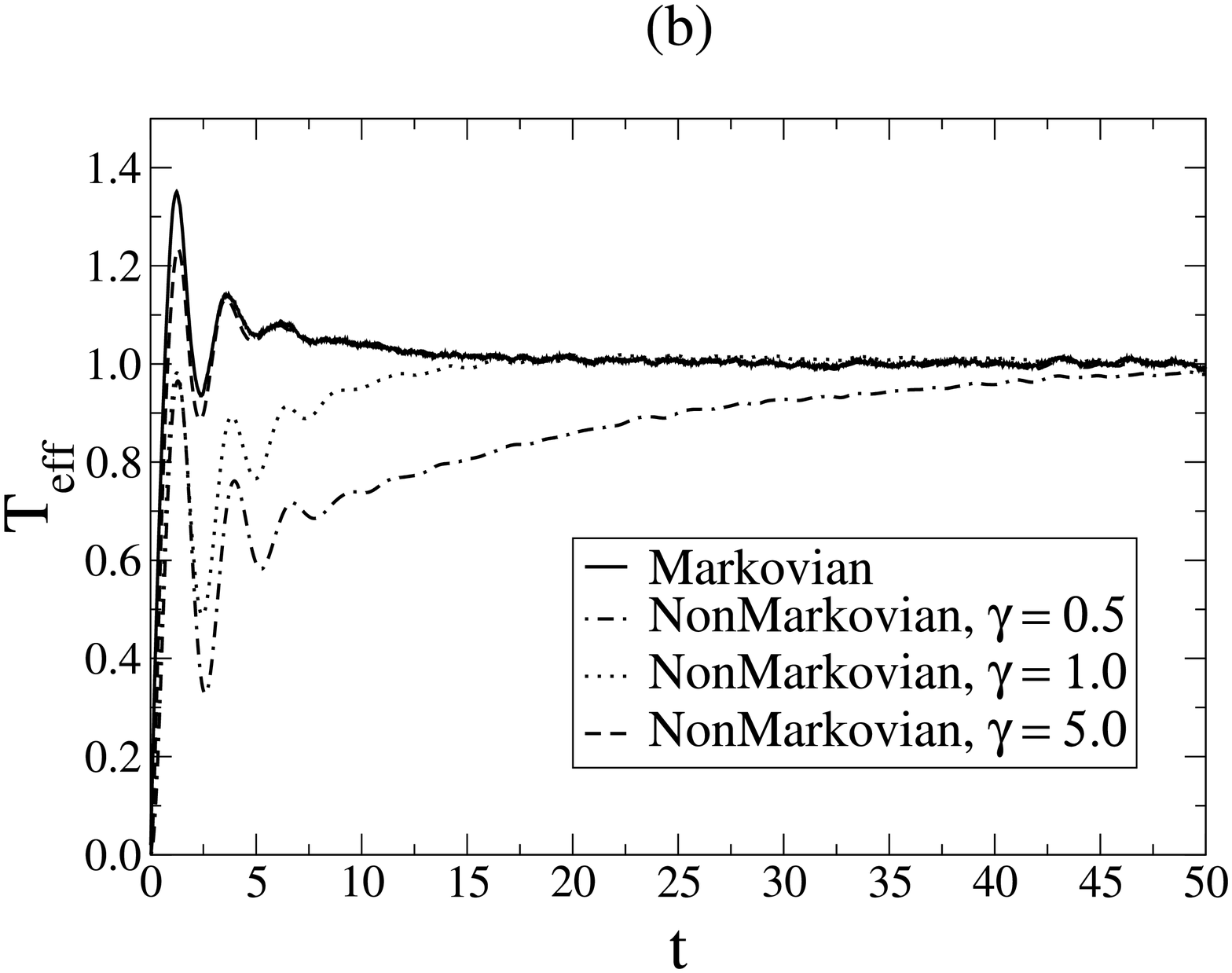,scale=0.265,angle=0}}
\vspace{0.5 cm}
\centerline{ \psfig{file=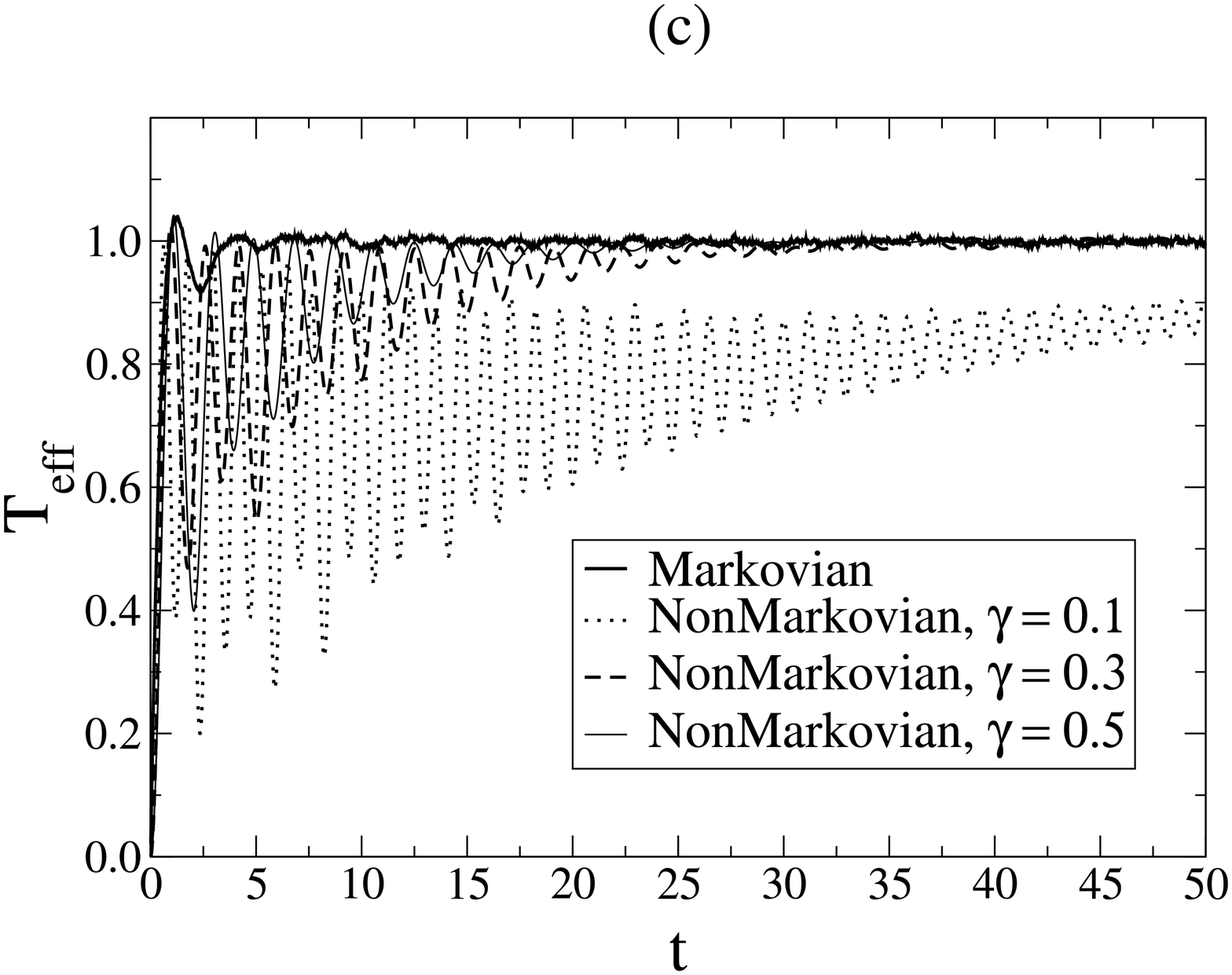,scale=0.265,angle=0}
  \psfig{file=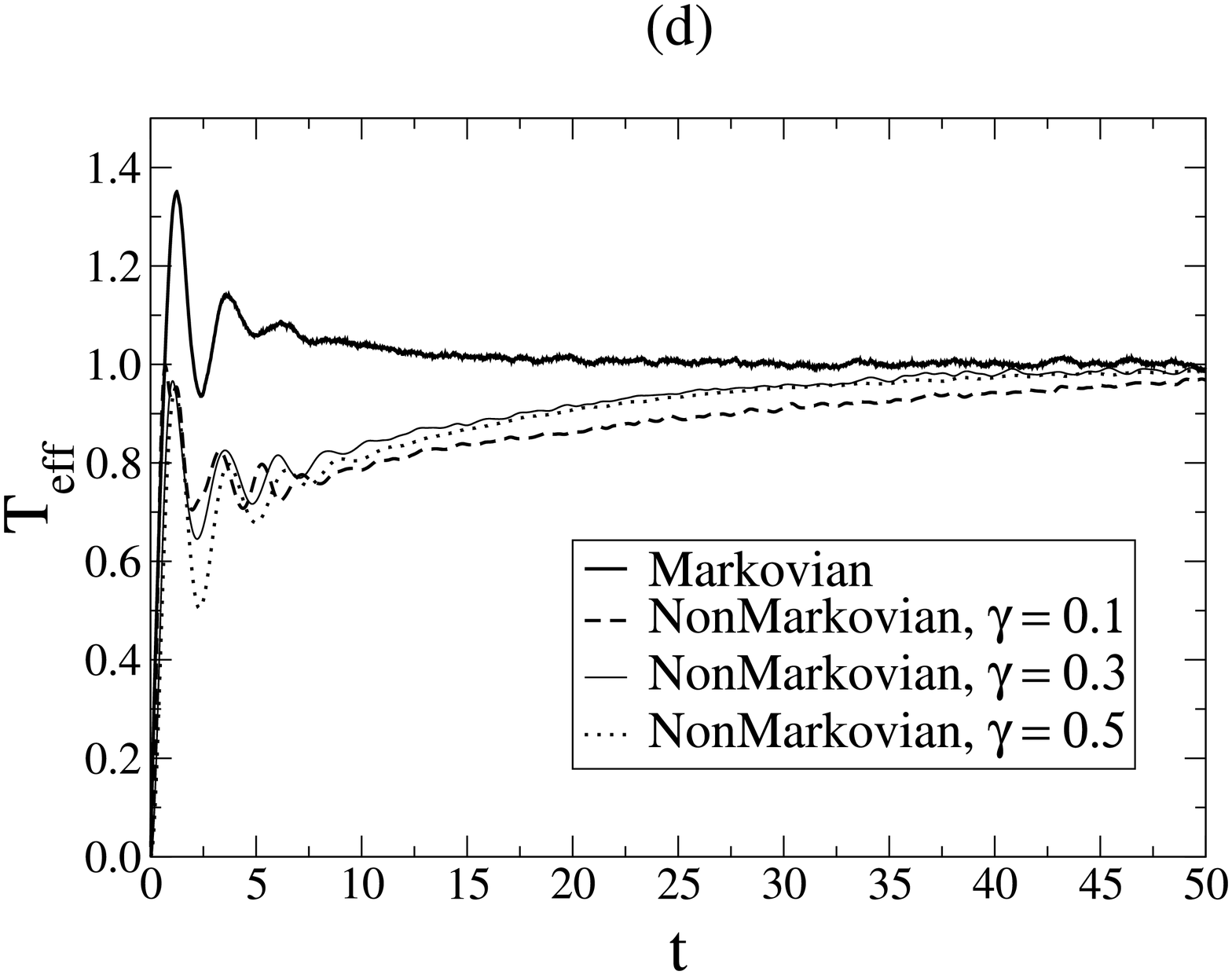,scale=0.265,angle=0}}
   \caption{\sf The effective temperature $T_{eff}$ for the (a) OU $n=0$ case,
     (b) OU $n=1$ case, (c) EDH $n=0$ case and (d) EDH $n=1$ case.}
   \label{fig9}
\end{figure}

In addition to the above results obtained for the ensemble average system
variable  $\phi$, it is also useful to determine how the memory effects
influence the thermalization time for the system when put in contact with the
thermal bath at some temperature $T$.  {}For this, let us define an effective
time dependent temperature for the system according to the equipartition of
kinetic energy,

\begin{equation}
T_{\rm eff} (t) = \langle \dot{\phi}^2(t) \rangle\;.
\label{Teff}
\end{equation}
The results for $T_{\rm eff}$ for the OU and EDH cases for the  additive and
multiplicative noise situations are shown in {}Fig. \ref{fig9},  where it is
also plotted the Markovian, local approximation  case, for comparison.

{}From the plots shown in {}Fig. \ref{fig9} we again see the overall behavior
seen in the previous plots. The larger is the relaxation time for the memory
kernels ({\it i.e.}, the smaller is the damping parameter $\gamma$ of the
memory kernels), 
the larger the dynamics takes to approach the Markovian approximated
one. We also see how memory effects  reflects in the thermalization of the
system. Larger bath relaxation times lead to a longer time for the system to
thermalize.  Typically, for comparable bath relaxation time scales, in the
additive case the system tends to thermalize faster than in the multiplicative
case. In Table \ref{tab2} we give the approximate time (in units of $1/m$) for
thermalization for all the cases studied above and where this behavior for the
time for thermalization in each of the types of dynamics analyzed can be
verified. Note that in all cases, with additive  or multiplicative noise, the
Markovian dynamics tend in general  to underestimate the time scale for
thermalization, when compared to the non-Markovian dynamics, except for large
$\gamma$ in the OU additive noise case, where the  thermalization in the
non-Markovian dynamics tends to be better approximated by the Markovian one.
As far the different dynamics are concerned, we observe from both the plots in
{}Fig. \ref{fig9} as from the results shown in Table \ref{tab2} for the
thermalization times, that the additive noise case
always tend to have a smaller thermalization time than the multiplicative
noise case. This is observed for both the Markovian and non-Markovian cases.

\begin{table}[pt]
{\begin{tabular}{@{}c|c|c|c|c@{}}  
\hline  
$\gamma$ & $\tau_{\rm
    markov \; add}$ & $\tau_{\rm OU \; add}$ &  $\tau_{\rm markov \; mult}$ &
$\tau_{\rm OU \; mult}$  \\ 
\colrule 
0.5  &   & 37  &     & 72 \\ 
1.0  & 4 & 18  & 27  & 68 \\ 5.0  &   & 4   &    & 38 \\ 
\hline 
\end{tabular}}
\hspace{0.5cm} {\begin{tabular}{@{}c|c|c|c|c@{}}  
\hline  
$\gamma$ &
    $\tau_{\rm markov \; add}$ & $\tau_{\rm EDH \; add}$ &  $\tau_{\rm markov
      \; mult}$ & $\tau_{\rm EDH \; mult}$  \\ 
\colrule 
0.1  &   & 152 &    & 126 \\ 
0.3  & 4 & 39  & 27 & 77 \\ 
0.5  &   & 33  &    & 74 \\ 
\hline 
\end{tabular}}
\caption{The approximate time for thermalization, in units of $1/m$,  for the
  Markovian and non-Markovian dynamics, determined when (\ref{Teff})
  approaches the temperature of the thermal bath.}
\label{tab2}
\end{table}

\subsection{Markovian and non-Markovian dynamics in terms of the
temperature of the thermal bath}

We now explore how the temperature of the thermal bath will influence the
results shown in the previous section. We again concentrate on the differences
between the Markovian and non-Markovian dynamics and the  thermalization
time. {}For this study we consider the cases with the highest values of
$\gamma$ considered in the previous subsection, which gives the best
comparison between the dynamics. Keeping the highest values of $\gamma$ allows
us to determine whether the comparison between  the two dynamics improves  or
worsens as the temperature is changed. 

\begin{figure}[htb]
 \centerline{ \psfig{file=fig10a.eps,scale=0.265,angle=0}
   \psfig{file=fig10b.eps,scale=0.265,angle=0}}
\vspace{0.95 cm}
\centerline{ \psfig{file=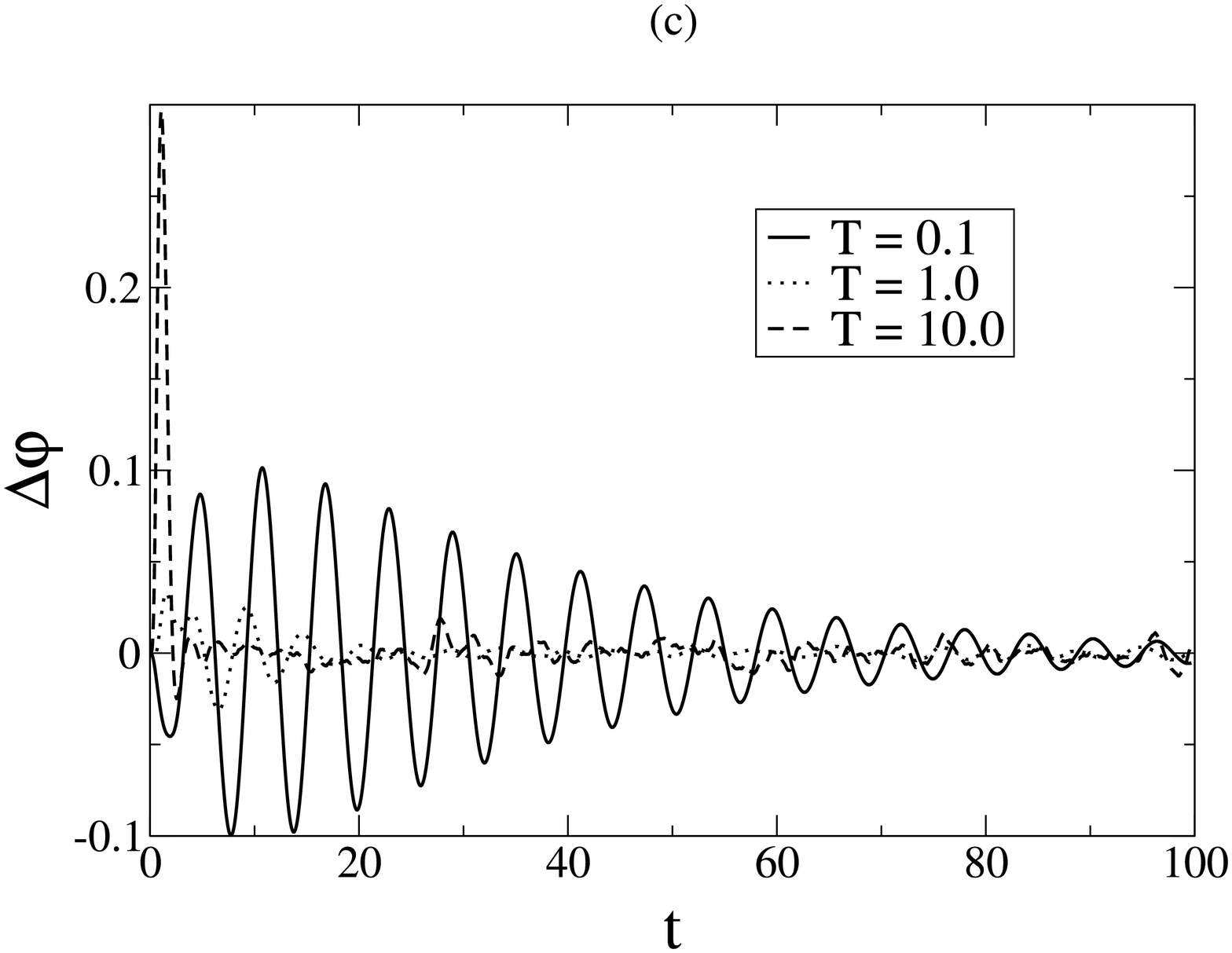,scale=0.265,angle=0}
  \psfig{file=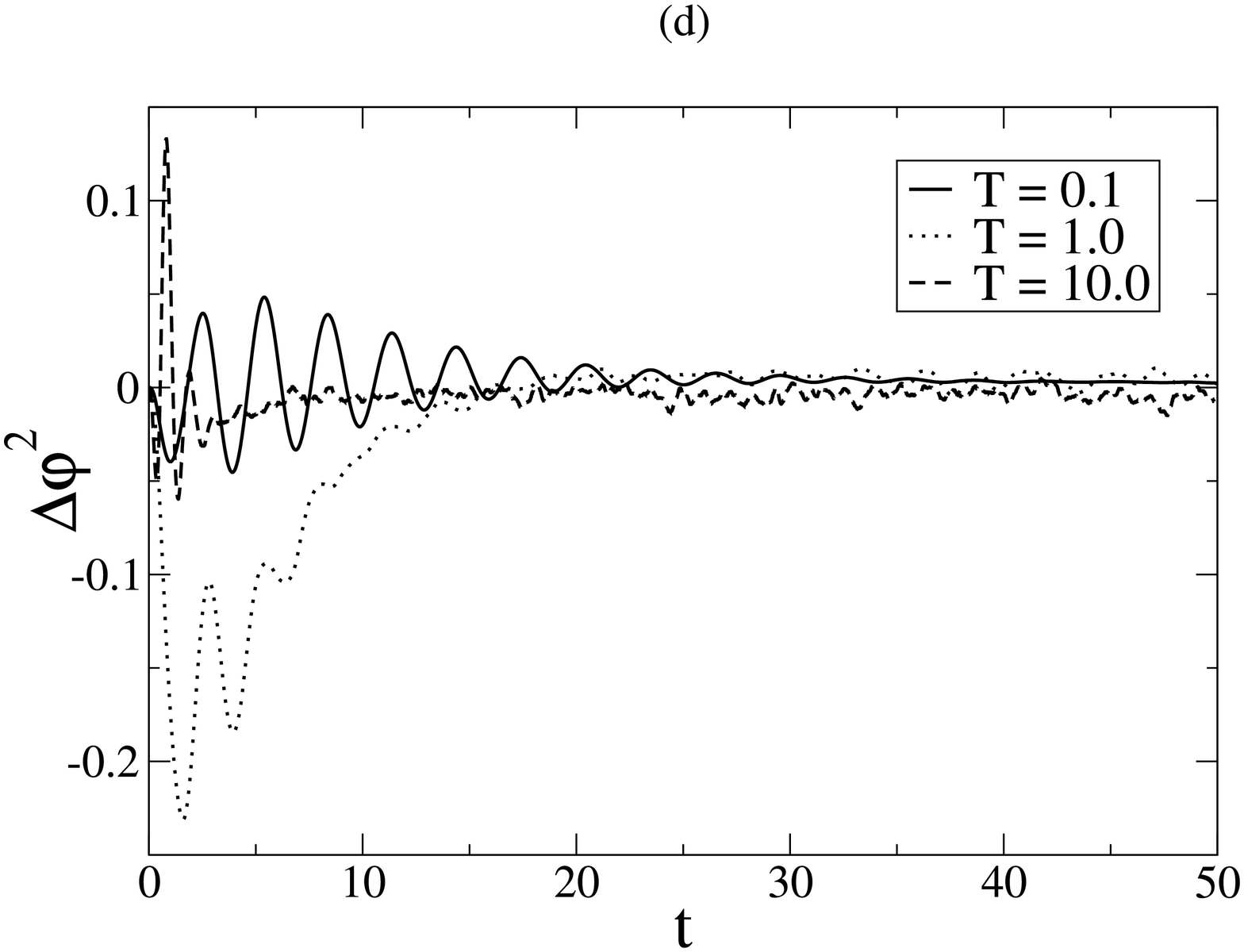,scale=0.265,angle=0}}
   \caption{\sf The results for $\Delta\phi$ and $\Delta \phi^2$ for the OU
     $n=0$ case, (a) and (b), respectively,  and for the OU $n=1$ case, (c) and
     (d), respectively.}
   \label{fig10}
\end{figure}

\begin{figure}[htb]
 \centerline{ \psfig{file=fig11a.eps,scale=0.265,angle=0}
   \psfig{file=fig11b.eps,scale=0.265,angle=0}}
\vspace{0.5 cm}
\centerline{ \psfig{file=fig11c.eps,scale=0.265,angle=0}
  \psfig{file=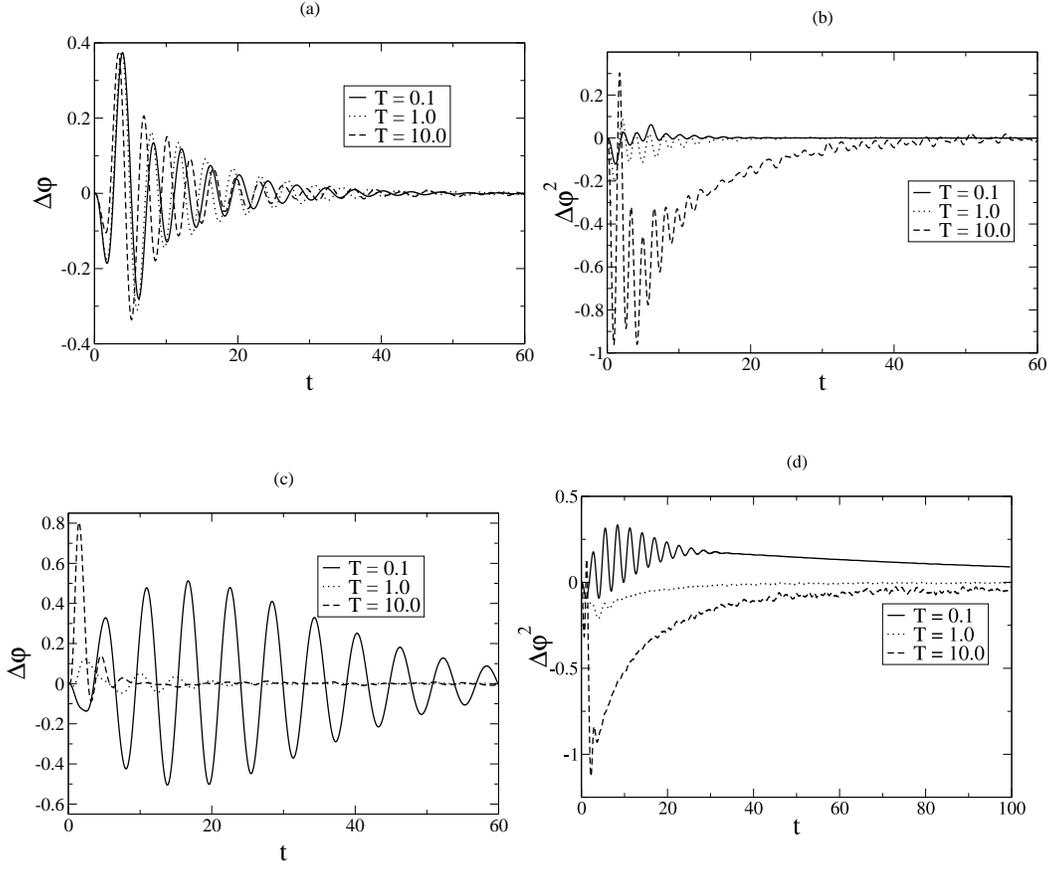,scale=0.265,angle=0}}
   \caption{\sf The results for $\Delta\phi$ and $\Delta \phi^2$ for the EDH
     $n=0$ case, (a) and (b), and for the EDH $n=1$ case, (c) and (d).}
   \label{fig11}
\end{figure}

The results for the differences $\Delta\phi$ and $\Delta \phi^2$ for the GLE
with OU kernel are shown in {}Fig. \ref{fig10}, while for the  GLE with EDH
kernel are shown in {}Fig. \ref{fig11}.  The time scale for which the
full non-Markovian dynamics approaches the respective Markovian dynamic
approximation,
in each of the cases studied here, are tabulated in Table \ref{tab3}.

\begin{table}[pt]
{\begin{tabular}{@{}c|c|c|c|c|c|c|c|c@{}}  
\hline  
$T$ & $\Delta
    \phi_{\rm OU \; add}$ & $\Delta \phi^2_{\rm OU \; add}$ &  $\Delta
    \phi_{\rm OU \; mult}$ & $\Delta \phi^2_{\rm OU \; mult}$  & $\Delta
    \phi_{\rm EDH \; add}$ & $\Delta \phi^2_{\rm EDH \; add}$ &  $\Delta
    \phi_{\rm EDH \; mult}$ & $\Delta \phi^2_{\rm EDH \; mult}$  \\ 
\colrule
    0.1  & 17  & 10  & 140  & 206 & 68 & 24  & 126  & 220 \\ 1.0  & 14  & 7
    & 27  & 12 & 53  &26  &23  & 67  \\ 10.0  & 13  & 6   & 6  & 7 & 42 & 49
    & 13 & 61 \\ 
\hline 
\end{tabular}}
\caption{The approximate time scale, in units of $1/m$, for the non-Markovian
  dynamics to approach the Markovian one, within a precision of $10^{-3}$ for
  the differences defined in (\ref{Deltas}) for different temperatures for the
  thermal bath. The parameter $\gamma$ considered is the largest one
  considered in Table \ref{tab1} in the OU ($\gamma =5.0$) and EDH cases
  ($\gamma=0.5$).}
\label{tab3}
\end{table}

{}From the results shown in Table \ref{tab3} we can see that higher
temperatures for the thermal bath give only minor improvements in the OU
additive noise case as regarding the approach of the non-Markovian dynamics to
the approximated Markovian one. But in the multiplicative (OU) case the
changes are much stronger, with $\langle \phi \rangle $ and 
$\langle \phi^2 \rangle$
approaching much faster to the Markovian dynamics the larger is the
temperature. In the EDH case, for the additive
noise, $\langle \phi \rangle $ tends to approach slower to the Markovian 
approximation and
in the multiplicative noise case the approach is much faster, 
compared to the additive one.
But the correlation  $\langle \phi^2 \rangle$ changes
much differently. In the EDH additive noise case the Markovian approximation
worsens as the temperature is increased, while in the multiplicative case the
Markovian approximation improves, but only slightly for high temperatures. 
Similar behavior to this will also be seen
below for the  thermalization times for each of the dynamics. We can also
observe from the results shown in the plots of {}Figs. \ref{fig10}  and
\ref{fig11} that for the multiplicative noise case, for both non-Markovian
dynamics studied, the memory effects are much stronger at lower temperatures
than at high temperatures and these effects last for a much longer time than
in the additive noise cases.

\begin{figure}[htb]
 \centerline{ \psfig{file=fig12a.eps,scale=0.265,angle=0}
  \hspace{0.1cm} \psfig{file=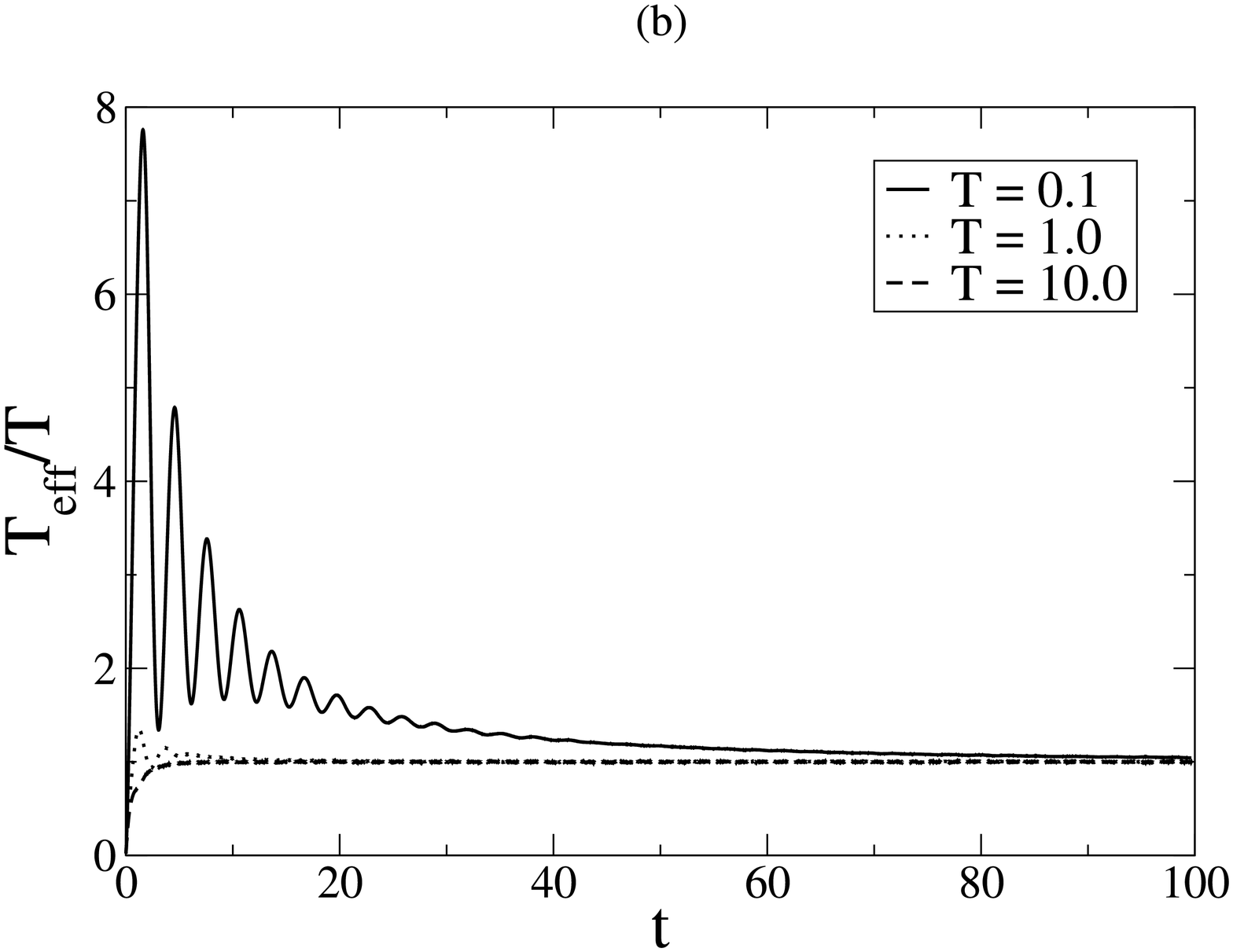,scale=0.265,angle=0}}
\vspace{0.55 cm}
\centerline{ \psfig{file=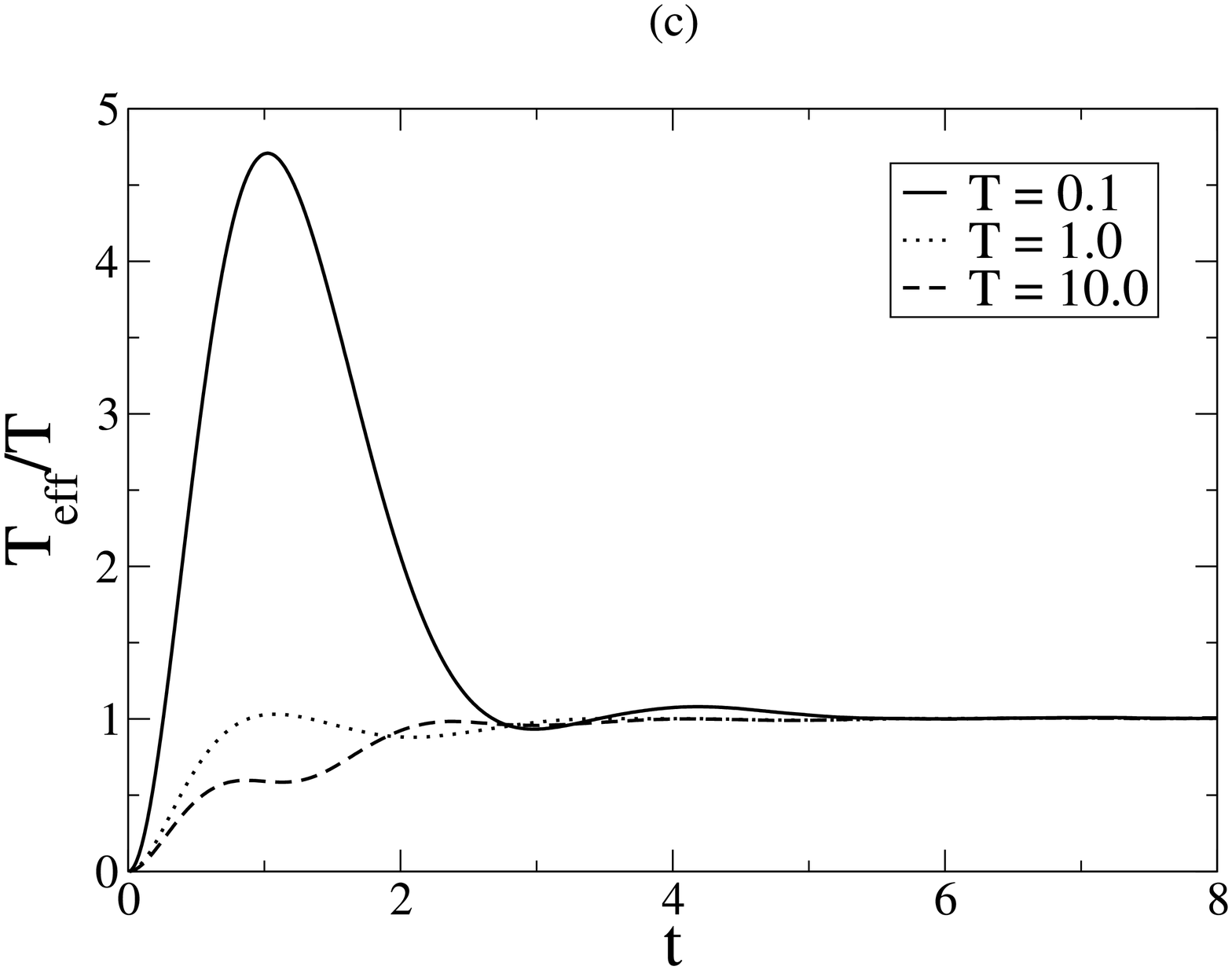,scale=0.265,angle=0}
 \hspace{0.1cm} \psfig{file=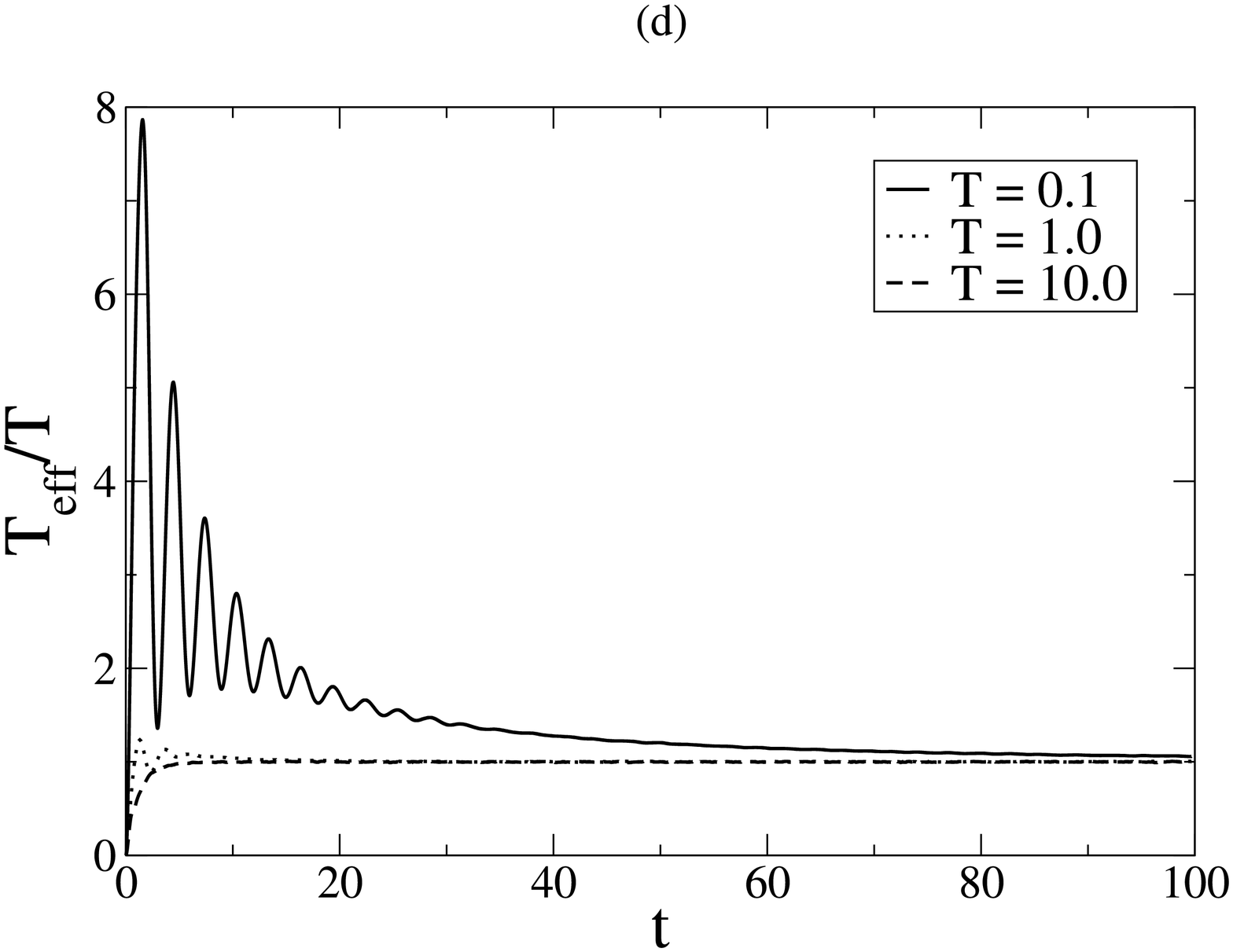,scale=0.265,angle=0}}
\vspace{0.55 cm}
 \centerline{
  \psfig{file=fig12e.eps,scale=0.265,angle=0}
  \hspace{0.1cm} \psfig{file=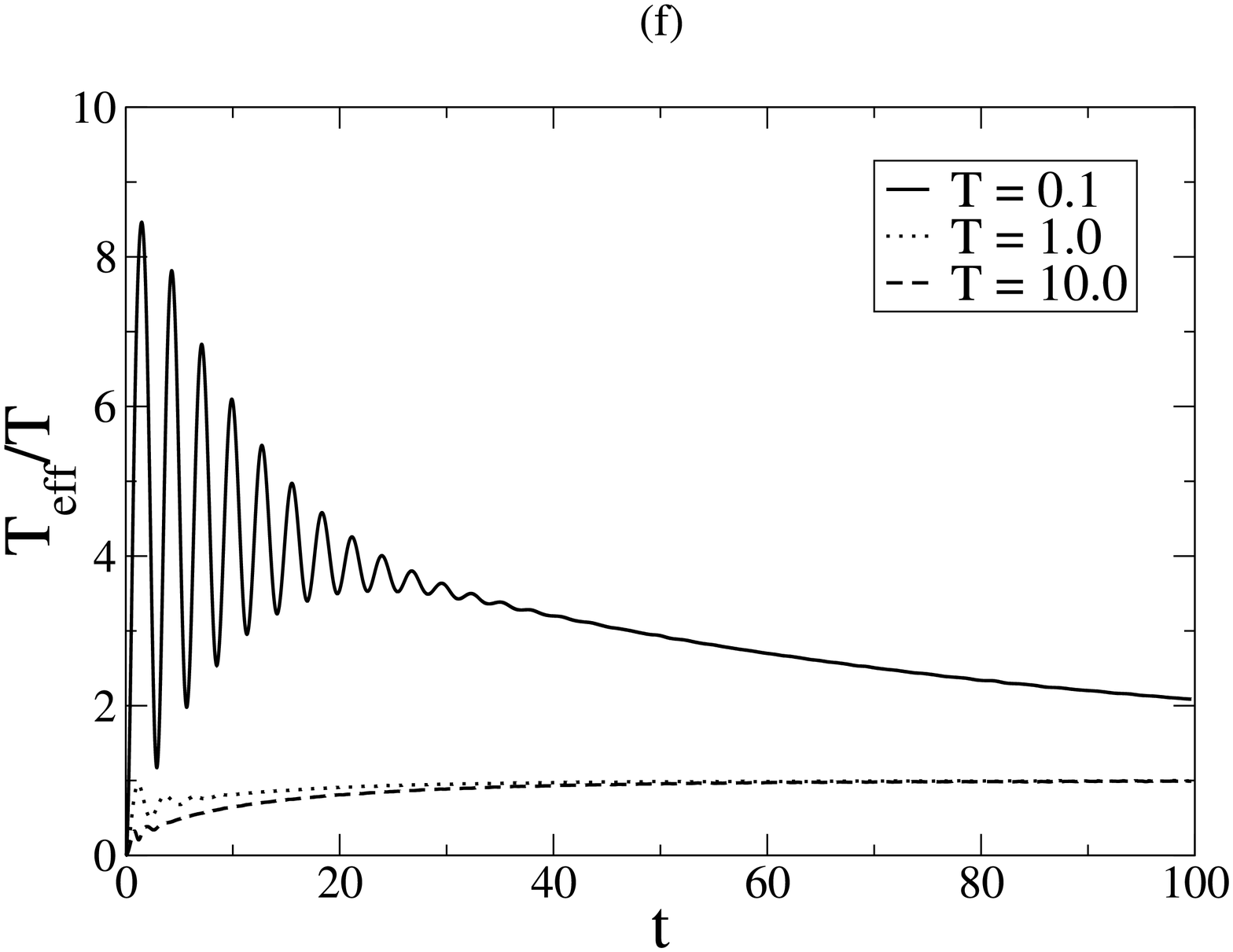,scale=0.265,angle=0}}
   \caption{\sf The results for the effective temperature as a function of
     time, when the temperature of the thermal bath is changed, for the cases
     of (a) Markovian additive noise, (b) Markovian multiplicative noise, (c)
     OU additive noise, (d) OU multiplicative noise, (e) EDH additive noise
     and (f) EDH multiplicative noise.}
   \label{fig12}
\end{figure}

We now study  how the thermalization time for each of the dynamics studied
here will change when the temperature is changed.  In {}Fig. \ref{fig12} we
show the plots for the various cases of dynamics studied here. The
thermalization times for each case are tabulated in Table \ref{tab4}.

\begin{table}[pt]
{\begin{tabular}{@{}c|c|c|c|c|c|c@{}}  
\hline 
$T$ & $\tau_{\rm
    markov \; add}$ & $\tau_{\rm markov \; mult}$ &  $\tau_{\rm OU \; add}$ &
$\tau_{\rm OU \; mult}$ & $\tau_{\rm EDH \; add}$ &  $\tau_{\rm EDH \; mult}$
\\ 
\colrule 0.1  & 7  &  123    & 6   &135 & 30 & 570 \\ 
1.0  & 4 & 27   & 4
& 38 & 33 & 74 \\ 10.0  & 6  &  9   & 7   & 15 & 48 & 65 \\ 
\hline 
\end{tabular}}
\caption{The approximate time for thermalization, in units of $1/m$,  for the
  Markovian and non-Markovian dynamics, determined when (\ref{Teff})
  approaches the temperature of the thermal bath.}
\label{tab4}
\end{table} 

{}From the results shown in the plots of {}Fig. \ref{fig12} and in
Table \ref{tab4}, we can clearly see that while the cases with OU noise the
thermalization times for the Markovian and non-Markovian dynamics are similar
(using the largest values of $\gamma$ considered in  Table \ref{tab1}),  the
non-Markovian dynamics with EDH noise, for both the additive and
multiplicative cases, produces thermalization  times considerably larger than
those of the Markovian cases.  As in the case of $\Delta \phi^2$ for the EDH
additive noise case seen in Table \ref{tab3}, we again see here the anomalous
behavior of the thermalization time increasing as the temperature is
increased in the EDH additive noise case. This is opposite to the behavior
seen in the OU and EDH noise cases with multiplicative noise.  Note also that
in the case closer to the Markovian dynamics, the OU additive noise case, 
there is almost no relevant difference in thermalization times as the
temperature is varied up to 2 orders of magnitude. {}For other values of
temperature of the thermal bath we tested, we found that these conclusions do
not change.

\subsection{The non-Markovian dynamics in terms of the
frequency of the thermal bath}

\begin{figure}[htb]
 \centerline{ \psfig{file=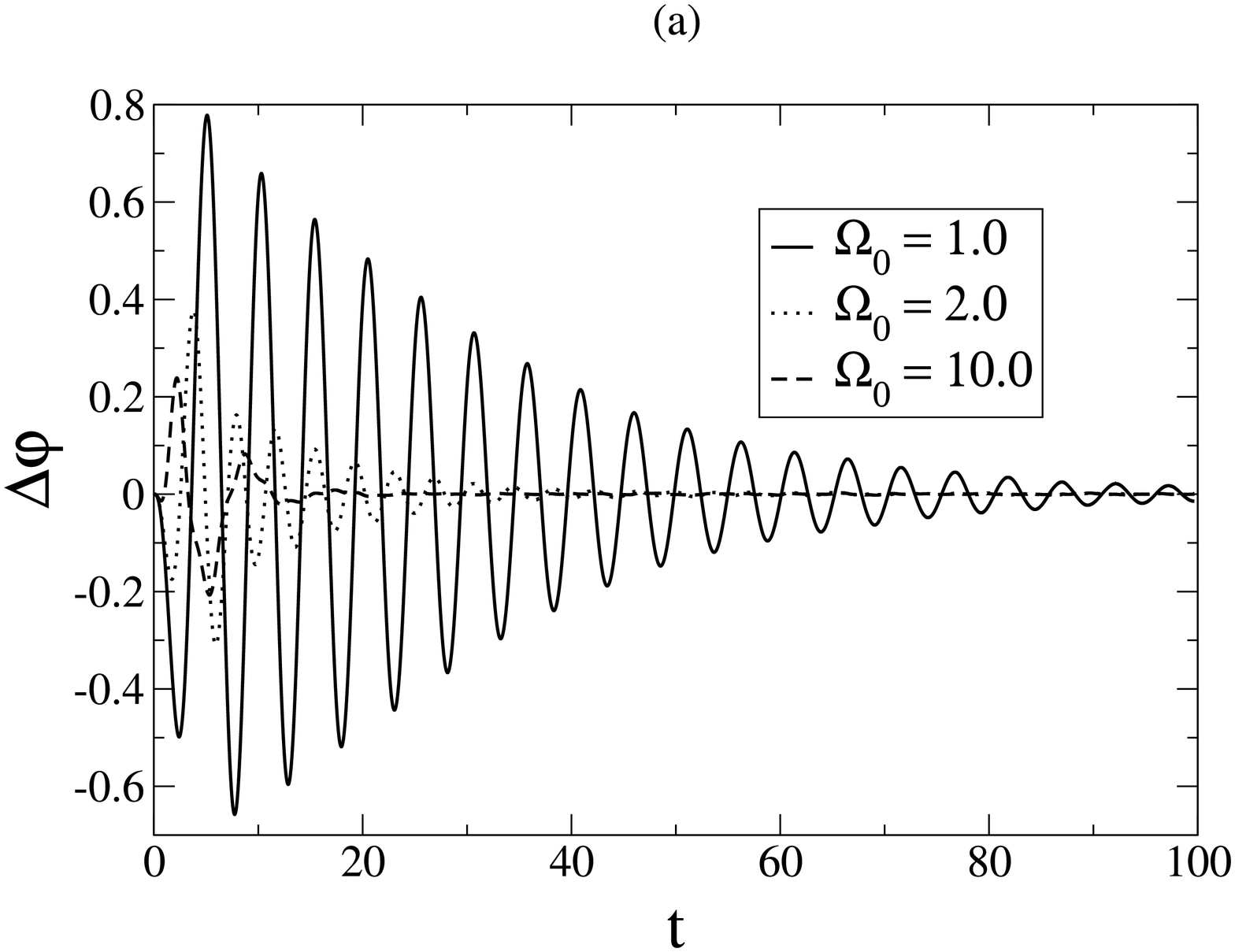,scale=0.265,angle=0}
   \psfig{file=fig13b.eps,scale=0.265,angle=0}}
\vspace{0.5 cm}
\centerline{ \psfig{file=fig13c.eps,scale=0.265,angle=0}
  \psfig{file=fig13d.eps,scale=0.265,angle=0}}
   \caption{\sf The results for $\Delta\phi$ and $\Delta \phi^2$, when the
     frequency of the thermal bath (in units of $m$) is changed, for the EDH
     $n=0$ case, plots (a) and (b), respectively,  and for the EDH $n=1$ case,
     plots (c) and (d), respectively.}
   \label{fig13}
\end{figure}

\begin{figure}[htb]
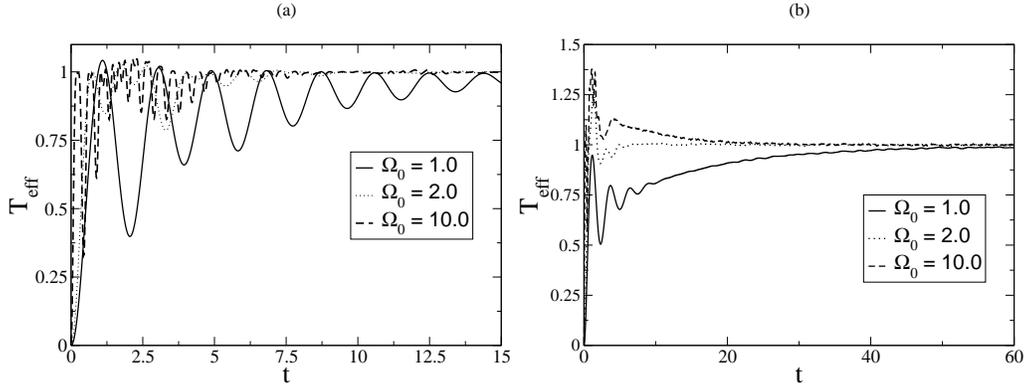

 \centerline{ \psfig{file=fig14a.eps,scale=0.265,angle=0}
   \psfig{file=fig14b.eps,scale=0.265,angle=0}}
   \caption{\sf The results for the effective temperature as a function of
     time, when the frequency of the thermal bath (in units of $m$) is
     changed, for the cases  of (a) EDH additive noise and (b) EDH
     multiplicative noise.}
   \label{fig14}
\end{figure}

We now investigate how the frequency of the thermal bath affects the  dynamics
of the system. Note that in this case only the GLE with EDH kernel is affected
(the Markovian and OU dynamics are independent of the frequency of the thermal
bath). As before, we can better interpret the results by analyzing the
differences (\ref{Deltas}) and the graph  of the effective temperature. {}From
them we obtain the time scale for the non-Markovian dynamics to approach the
Markovian one and we can better estimate the differences in thermalization
times (if any) in both cases. The results for this case, when the frequency of
thermal bath is varied, are shown in {}Fig. \ref{fig13} for the differences
(\ref{Deltas}), while the behavior of the effective temperature, when the 
frequency of the bath is changed, is shown in {}Fig. \ref{fig14}. 
In all cases the
frequency  $\Omega_0$ is in units of the frequency of the system, $m$, and the
time is in units of $1/m$. As in the previous analysis for the EDH case, 
we consider $\gamma=0.5$,
and all other parameters kept fixed at the values as given before, except
$\Omega_0$.  The time scales for the non-Markovian and Markovian dynamics to
approach  to each other are given in Table \ref{tab5}, while the time for
thermalization for each case is given in Table \ref{tab6}.

\begin{table}[pt]
{\begin{tabular}{@{}c|c|c|c|c@{}}  
\hline 
$\Omega_0/m$ & $\Delta
    \phi_{\rm EDH \; add}$ & $\Delta \phi^2_{\rm EDH \; add}$ &  $\Delta
    \phi_{\rm EDH \; mult}$ & $\Delta \phi^2_{\rm EDH \; mult}$  \\ 
\colrule
    1.0  & 53  & 26  & 23  & 67  \\ 2.0  & 24  & 8   & 25  & 38  \\ 10.0 & 22
    & 8   & 25  & 33   \\ \hline 
\end{tabular}}
\caption{The approximate time scale, in units of $1/m$, for the non-Markovian
  dynamics to approach the Markovian one, within a precision of $10^{-3}$ for
  the differences defined in (\ref{Deltas}) for different frequencies of the
  thermal bath. The parameter $\gamma$ considered is the largest one
  considered in Table \ref{tab1} in the EDH cases ($\gamma=0.5$).}
\label{tab5}
\end{table}

\begin{table}[pt]
{\begin{tabular}{@{}c|c|c|c|c@{}}  
\hline 
$\Omega_0/m$ & $\tau_{\rm
    markov \; add}$  & $\tau_{\rm markov\; mult}$  &  $\tau_{\rm EDH \; add}$
& $\tau_{\rm EDH \; mult}$  \\ 
\colrule 
1.0  &   &    & 33  &  74  \\ 
2.0  & 4
& 27 & 13  &   8  \\ 
10.0 &   &    & 9   & 35   \\ 
\hline 
\end{tabular}}
\caption{The approximate time for thermalization, in units of $1/m$,  for
  different frequencies of the thermal bath in the in the Markovian
  approximation and in the  EDH additive and multiplicative noise cases.}
\label{tab6}
\end{table}

{}From the results seen in {}Fig. \ref{fig13} and Table \ref{tab5} we can note
that the smaller is the frequency of the thermal bath the larger  it takes for
both $\langle \phi \rangle$ and $\langle \phi^2 \rangle$  in the non-Markovian
case to approach the Markovian approximation.  As the frequency of the bath is
increased beyond the frequency of the system, the time scale for the
non-Markovian dynamics to approach the Markovian one tends to  decrease, but
it also rapidly reaches a point where larger frequencies for the bath would not
make the Markovian approximation to improve too much.
 
The results are seen to be much different again when the thermalization of the
system is studied. {}From the thermalization plots seen in {}Fig. \ref{fig14} 
and the thermalization times given in Table \ref{tab6}, we see that
smaller frequencies for the bath, compared to the system's frequency, lead to
much higher thermalization times for the non-Markovian approximation  compared
to the Markovian one. The thermalization time for the non-Markovian dynamics
is seen to decrease with an increase of $\Omega_0$, except the multiplicative
noise case, where above a certain point $\Omega_0 \gtrsim m$, the
thermalization time starts to increase again. Though the importance  of the
memory kernel is expected to be less important the larger is $\Omega_0$,
which makes it fast oscillate, the behavior in the multiplicative noise is
probably far less simple, because the higher are the nonlinearities  in that
case, which are brought by the system dependent dissipation and noise.

\section{Conclusions}

In this work we have analyzed in details the differences between the dynamics
of a system when treating it in terms of its full non-Markovian equation of
motion and when expressing it in terms of its Markovian, or local approximated
form. Having set the appropriate description for the non-Markovian equations,
we have then studied the applicability of the local approximation for these
equations. We here have concentrated in two forms for the non-Markovian memory
kernel: The OU and EDH cases and we have analyzed the cases of additive and
multiplicative noises in both cases. We have seen that in general, for most of
the parameters in both cases, the local  approximation is far from being to
represent a good description of the dynamics. Obviously, since these are all
dissipative systems, we expect the two dynamics, non-Markovian and
Markovian, to tend to each other asymptotically. We have then analyzed how
long it takes for each of the non-Markovian dynamics studied to tend to the
Markovian ones. We have  analyzed both the time for the system variable,
$\langle \phi \rangle $, as well for the equal time correlation, 
$\langle \phi^2 \rangle$. We
have also analyzed the thermalization time for each of the dynamics by
studying the  behavior of the correlation function $\langle \dot{\phi}^2
\rangle$, which, according to the equipartition theorem, can be associated to
the temperature of equilibration of the system. 

Each of the dynamics were investigated by changing the main parameters
characterizing the thermal bath: The damping term for the memory kernels,
$\gamma$, the temperature $T$ and the frequency $\Omega_0$  of the thermal
bath. The parameters of the system were kept fixed for convenience as
well the magnitude of the dissipation, $\eta$, which is a linear parameter
entering in all the dynamics and that was kept fixed at the point where all
the dynamics were initially underdamped. By increasing $\gamma$, the memory
kernels are damped faster and the Markovian approximation tends to be better
as expected. As $\gamma$ is varied, besides the expected behavior of the
non-Markovian dynamics to approach faster the Markovian one the larger is the
$\gamma$,  it is also observed that for the same value for the memory kernel
damping term, in general the cases with EDH non-Markovian dynamics  tend to
approach faster to the Markovian dynamics than the OU cases, as far the
thermalization times in each of the dynamics are concerned. It is also
observed that in general the Markovian dynamics tend to  overestimate the time
for thermalization as compared to the time it takes in the non-Markovian
cases. The thermalization times in the studied dynamics also tend to be larger
in the multiplicative noise cases than in the additive ones.

When analyzing the behavior of each of the dynamics when varying the
temperature and frequency of the thermal bath, we have fixed the value of
$\gamma$ and then investigated each of the time scales for non-Markovian
dynamics for $\langle \phi \rangle $ and $\langle \phi^2 \rangle$ 
to approach the corresponding
Markovian approximation. By increasing the temperature it is observed that the
Markovian approximation tends to improve, except in the case of the dynamics
of the correlation $\langle \phi^2 \rangle$ in the EDH case with additive
noise, where the Markovian approximation tends to worsen and in the OU
additive noise case, where the dynamics is weakly dependent on the temperature
of the thermal bath. These results seem to indicate that the effective
dissipation in the other dynamics is dependent on the temperature of the
thermal bath, in particular in the multiplicative noise cases, where the
effective dissipation seems to increase much faster with the 
temperature than in the
additive noise cases. The thermalization times shown in Table \ref{tab4} are
also indicative of this behavior.  {}Finally, in the study of how the
frequency of the thermal bath affects the dynamics (in the EDH memory kernel
cases), we have seen that as the frequency of the thermal bath is increased, 
the Markovian approximation seems to improve till some frequency close to the
system's frequency.  Above that value there is little improvement. We have
also verified that for frequencies of the thermal bath much below the system's
one, the Markovian approximation worsens considerably.

A few generic results can also be drawn from the analysis of all cases studied
here. In particular, we can note from the obtained results that either the
local approximation underestimates the effective dissipation seen in the
non-Markovian dynamics,  or overestimates it in most of the regions of
parameters. The local (Markovian)  approximation for the dynamics tends to be
better at larger values of the  bath damping term $\gamma$ and for larger
values of the frequency $\Omega_0$  and temperature $T$  of the thermal bath
(except for the correlation $\langle \phi^2 \rangle$ in the case of EDH 
memory kernel with additive
noise).  The difference between the dynamics is larger at short times,
exactly as expected because of the finite memory times for the non-Markovian
equations. The different simulations we have performed with different bath
parameters allowed us to estimate the approximate time scales when the
Markovian approximation may become an appropriate description of the
dynamics. In general, this time scale is much larger in the multiplicative
noise cases than in the additive noise ones.  Also, given the specific
differences seen in each of the dynamics when  varying e.g. either $\gamma$,
$T$ or $\Omega_0$, by looking at these  differences when changing the
properties of the thermal bath  may be a useful way to discriminate possible
stochastic phenomena in nature and to tell whether they can be dominated by
additive or multiplicative noises.

{}Finally, we should point out the possible connection and relevance 
of our studies
with those related to the dynamics of field theory models. It has been shown
\cite{GR,BMR} that non-Markovian kernels of the same form as studied here can
also appear in the studies of the effective dynamics of an order parameter in
field theories and in cosmology in general. Since the studies of
dissipative processes in those applications are related to  time non-local
terms in the effective evolution equation of the system, for example, for a
background scalar field describing the system, or an order parameter for a
phase  transition problem, we expect our results to be of relevance for
understanding the relevant dynamics in those situations as well. In
particular, our studies may be useful to clarify the applicability or not of
approximating the dynamics in those problems  as local ones, as usually it is
considered to be the case there.   The results we have obtained here shows
that, in many cases, the local approximation is not a reliable description of
the true non-Markovian dynamics. The difference can be very  large at short
times and continue to be for long time scales,  with memory effects making a
strong contribution  for the dynamics.  This may have strong consequences, for
example, when studying thermalization and equilibration times in phase
transition problems, or in the problem  of the production of particles and
radiation in cosmology.  In the dynamics of some systems in contact with a
thermal bath, as we have seen in the studies performed in this work, the usual
local Langevin equation typically underestimates the thermalization with
respect to the true dynamics, indicating that the use of local approximated
forms for the study of the dynamics can be unappropriated and even lead to
erroneous results as regarding to the system's equilibration and
thermalization time scales.


\acknowledgments The authors would like to thank Funda\c{c}\~ao de Amparo \`a
Pesquisa do Estado do Rio de Janeiro (FAPERJ) and Conselho Nacional de
Ci\^encia e Tecnologia (CNPq) for financial support.



\begin{thebibliography}{99}

\bibitem{reviews} N. G.  van Kampen, {\it Stochastic processes  in physics and
  chemistry}, 2nd  Ed. (North-Holland,  Amsterdam, 1992); H.  S. Wio,  {\it An
  introduction   to  stochastic   processes  and   nonequilibrium  statistical
  physics},  Series  on Advances  in  Statistical  Mechanics,  Vol. 10  (World
  Scientific, Singapore, 1994).

\bibitem{critical} K. Kawasaki, in {\it Phase Transitions and Critical
  Phenomena}, Vol. 2, Eds. C. Domb and M. S. Green (Academic, New York, 1976).

\bibitem{critical2}   P. C. Hohenberg and B. I. Halperin, Rev. Mod. Phys. {\bf
  49}, 435 (1977).

\bibitem{CH} J. W. Cahn and J. C. Hilliard, J. Chem. Phys. {\bf 28}, 258
  (1958).
%
\bibitem{Jou1} D.~Jou, J.~Casas-V\'azquez and G.~Lebon, Rep. Prog. Phys. {\bf
  51},1105 (1988); {\it ibid.} {\bf 62}, 1035 (1999).
%
\bibitem{KKR}T. Koide, G.  Krein and R. O. Ramos, Phys.  Lett.  {\bf B636}, 96
  (2006).
%
\bibitem{oscillators}G. W. Ford, M. Kac and P. Mazur, J. Math. Phys. {\bf 6},
  504 (1965).
%
\bibitem{weiss}U. Weiss, {\it Quantum Dissipative Systems} (World Scientific,
  Singapore, 1999).
%
\bibitem{caldeira} A. O. Caldeira and A. J. Leggett, Phys. Rev. Lett. {\bf
  46}, 211 (1981); Ann.   Phys. (N.Y.) {\bf 149}, 374 (1983);  {\it ibid.}
  {\bf 153}, 445 (1984).
%
\bibitem{GR}M. Gleiser and R. O. Ramos, Phys. Rev. D {\bf 50}, 2441 (1994).
  A. Berera, M. Gleiser and R. O. Ramos, Phys.  Rev. D {\bf 58}, 123508
  (1998).
%
\bibitem{BMR}A. Berera, I. G. Moss and R. O. Ramos,  Rep. Prog. Phys. {\bf
  72}, 026901 (2009).
%
\bibitem{ingold} G.-L. Ingold, in {\it Quantum Transport and Dissipation},
  Chap. 4, Eds.  T. Dittrich, P. H\"anggi, G.-L. Ingold, B. Kramer,
  G. Sch\"on and W. Zwerger (Wiley-VCH, Weinheim, 1998).
%
\bibitem{hanggi}P. H\"anggi and P. Jung, Adv. Chem. Phys. {\bf 89}, 239
  (1995).
%
\bibitem{OUref}G. E. Uhlenbeck and L. S. Ornstein, Phys. Rev. {\bf 36},  823
  (1930).

\bibitem{EDHref}L. Schimansky-Geier and C. Z\"ulicke, Z. Phys. B: Condens.
  Matter {\bf 79}, 451 (1990); R. Bartussek, P. Hanggi, B. Lindner and
  L. S. Geier, Physica. D \textbf{109}, 17 (1997).

\bibitem{Luczka}J. Lucza, Chaos {\bf 15}, 026107 (2005).
%
\bibitem{CPC2008} R. L. S. Farias, R. O. Ramos and L. A. da Silva,
  Comp. Phys. Comm. {\bf 180}, 574 (2009).

%
\bibitem{bao1} K. L\"u and J.-D. Bao, Phys. Rev. E {\bf 72}, 067701 (2005).
%
\bibitem{bao2} J.-D. Bao, Y.-L. Song, Q. Ji and Y.-Z. Zhuo, Phys. Rev. E {\bf
  72}, 011113 (2005).


\end{thebibliography}
\end{document}